\documentclass[10pt,conference]{IEEEtran}
 \IEEEoverridecommandlockouts

\usepackage{subcaption}
\usepackage{graphicx}
\usepackage{graphbox}
\usepackage{hyperref}
\usepackage{setspace}
\usepackage{wrapfig}
\usepackage{xspace}

\usepackage[colorinlistoftodos,prependcaption,textsize=tiny]{todonotes}
\usepackage{verbatim}

\usepackage{graphicx}
\usepackage{color}
\usepackage{url}
\usepackage{hyperref}
\usepackage{listings}
\usepackage{csquotes}
\usepackage[normalem]{ulem}
\usepackage{setspace}
\usepackage{algorithm}
\usepackage[most]{tcolorbox}
\usepackage[noend]{algpseudocode}

\usepackage{amssymb}
\usepackage{xparse}
\usepackage{mathtools}
\usepackage{multirow}
\usepackage{datapie}
\usepackage{adjustbox}
\usepackage{threeparttable} 

\def\BibTeX{{\rm B\kern-.05em{\sc i\kern-.025em b}\kern-.08emT\kern-.1667em\lower.7ex\hbox{E}\kern-.125emX}}

\newcommand{\mybox}[1]{\begin{tcolorbox}[enhanced, frame hidden, boxsep=0pt]\emph{#1}\end{tcolorbox}}

\newcommand{\modifycr}[1]{{\color{blue}{#1}}}

\newcommand{\modify}[1]{{\color{blue}{#1}}}

\newcommand{\modifydata}[1]{{{#1}}}

\newcommand{\modifyshin}[1]{{{#1}}}

\newcommand{\devcomment}[1]{\begin{tcolorbox}[enhanced, frame hidden, boxsep=0pt]\emph{#1}\end{tcolorbox}}

\usepackage{eqparbox}
\newdimen{\algindent}
\algnewcommand\LeftComment[2]{%
\hspace{#1\algindent}$\triangleright$ #2 \hfill %
}

\usepackage{listings}
\usepackage{xcolor}

\definecolor{codegreen}{rgb}{0,0.6,0}
\definecolor{codegray}{rgb}{0.5,0.5,0.5}
\definecolor{uncoveredgray}{rgb}{0.8,0.8,0.8}
\definecolor{codepurple}{rgb}{0.58,0,0.82}
\definecolor{backcolour}{rgb}{0.99,0.80,0.87}

\lstdefinestyle{mystyle}{
    backgroundcolor=\color{backcolour},   
    commentstyle=\color{codegreen},
    keywordstyle=\color{magenta},
    numberstyle=\tiny\color{codegray},
    stringstyle=\color{codepurple},
    basicstyle=\ttfamily\footnotesize,
    breakatwhitespace=false,         
    breaklines=true,                 
    captionpos=b,                    
    keepspaces=true,                 
    numbers=left,                    
    numbersep=5pt,                  
    showspaces=false,                
    showstringspaces=false,
    showtabs=false,                  
    tabsize=2
}

\lstdefinestyle{demo-code}{
    backgroundcolor=\color{white},   
    commentstyle=\color{codegreen},
    keywordstyle=\color{magenta},
    numberstyle=\tiny\color{codegray},
    stringstyle=\color{codepurple},
    breakatwhitespace=false,         
    breaklines=true,                 
    captionpos=b,                    
    keepspaces=true,                 
    numbers=left,                    
    numbersep=5pt,                  
    showspaces=false,                
    showstringspaces=false,
    showtabs=false,                  
    tabsize=2
}

\lstdefinestyle{no_backcolor}{
    commentstyle=\color{codegreen},
    keywordstyle=\color{magenta},
    numberstyle=\tiny\color{codegray},
    stringstyle=\color{codepurple},
    breakatwhitespace=false,
    breaklines=true,         
    captionpos=t,    
    keepspaces=true,
    basicstyle=\ttfamily\small,
    numbers=left,                    
    showspaces=false,                
    showstringspaces=false,
    showtabs=false,                  
    tabsize=2,
    numbersep=0,
}

\newcommand{\uncovered}{\makebox[0pt][l]{\color{uncoveredgray}\rule[-4pt]{1.0\linewidth}{13pt}}}
\newcommand{\simulee}[1]{\textit{Simulee}}
\newcommand{\gklee}[1]{\textit{GKLEE}}
\newcommand{\curd}[1]{\textit{CURD}}
\newcommand{\barracuda}[1]{\textit{BARRACUDA}}
\newcommand{\eraser}[1]{\textit{Eraser}}
\newcommand{\llvm}[1]{\textit{llvm}}
\newcommand{\codeIn}[1]{\texttt{#1}}
\newcommand{\remove}[1]{}
\newcommand{\git}[1]{\textit{GitHub}}
\newcommand{\junit}[1]{\textit{JUnit}}
\newcommand{\Always}[1]{\textit{Test failure}}
\newcommand{\always}[1]{\textit{test failure}}
\newcommand{\Sometime}[1]{\textit{Flaky test}}
\newcommand{\sometime}[1]{\textit{flaky test}}
\newcommand{\Crash}[1]{\textit{Crash}}
\newcommand{\crash}[1]{\textit{crash}}
\newcommand{\Perform}[1]{\textit{Inferior performance}}
\newcommand{\perform}[1]{\textit{inferior performance}}
\newcommand{\Improper}[1]{\textit{Improper resource management}}
\newcommand{\improper}[1]{\textit{improper resource management}}
\newcommand{\Non}[1]{\textit{Non-optimal implementation}}
\newcommand{\non}[1]{\textit{non-optimal implementation}}
\newcommand{\cross}[1]{\textit{generic error}}
\newcommand{\Cross}[1]{\textit{Generic error}}
\newcommand{\Port}[1]{\textit{Poor portability}}
\newcommand{\Sync}[1]{\textit{Improper synchronization}}

\newcommand{\approach}[1]{\textsc{CDFuzz}}

\newcommand{\port}[1]{\textit{poor portability}}

\newcommand{\torder}[1]{\textit{Torder List}}

\newcommand{\vo}[1]{\textit{visit\_order}}
\newcommand{\ti}[1]{\textit{thread\_id}}
\newcommand{\ac}[1]{\textit{action}}

\newcommand{\afl}{AFL}
\newcommand{\aflpp}{AFL++}
\newcommand{\angora}{Angora}
\newcommand{\fairfuzz}{\textsc{FairFuzz}\xspace}
\newcommand{\mopt}{\textsc{Mopt}\xspace}
\newcommand{\qsym}{\textsc{Qsym}\xspace}
\newcommand{\meuzz}{\textsc{Meuzz}\xspace}
\newcommand{\pangolin}{\textsc{Pangolin}\xspace}
\newcommand{\redqueen}{\textsc{Redqueen}\xspace}

\newcommand{\afldict}{AFL$_{Dict}$}

\newcommand{\branch}{branch condition}
\newcommand{\branches}{branch conditions}
\newcommand{\original}{\textit{$Orig$}}
\newcommand{\dictionary}{\textit{$Dict$}}
\newcommand{\assistfuzzingspaper}{Assisting Exploration Strategies}

\newcommand{\assistfuzzings}{assisting exploration strategies}

\newcommand{\dictionarybased}{dictionary strategy}
\newcommand{\Dictionarybased}{Dictionary Strategy}
\newcommand{\inputstate}{input-to-state correspondence strategy}
\newcommand{\Inputstate}{Input-to-State Correspondence Strategy}
\newcommand{\gradient}{gradient-based strategy}

\newcommand{\allconstraints}{constraint-solver-based strategies}
\newcommand{\Allconstraints}{Constraint-Solver-based Strategies}
\newcommand{\hybrid}{SMT-solver-based strategy}

\newcommand{\fuzzingdriver}{FuzzingDriver}
\newcommand{\jumpequal}{\textit{Jump Equal instruction}}
\newcommand{\jumpnoequal}{\textit{Jump not Equal instruction}}
\newcommand{\equal}{\textit{equality constraint}}
\newcommand{\nonequal}{\textit{non-equality constraint}}

\newcommand{\totalrun}{five}
\newcommand{\totalfuzzer}{nine}
\newcommand{\totalcrash}{37}
\newcommand{\totalfixed}{seven}
\newcommand{\totalconfirmed}{nine}
\newcommand{\totalbench}{\modifydata{21}}

\newcommand{\unconfirmed}[1]{\textit{reported}}
\newcommand{\confirmed}[1]{\textit{confirmed}}
\newcommand{\pendingfix}[1]{\textit{confirmed}}
\newcommand{\fixed}[1]{\textit{confirmed and fixed}}

\newcommand{\Uouv}{Use-of-uninitialized-value}

\newcommand{\El}{Infinite loop}

\newcommand{\Ml}{Memory leaked}

\newcommand{\Sf}{Segmentation fault}

\newcommand{\Astb}{Allocation-size-too-big}

\newcommand{\Hbo}{Heap-buffer-overflow}
\newcommand{\hbo}{heap-buffer-overflow}
\newcommand{\Oom}{Out-of-memory}

\newcommand{\Af}{Assertion failure}
\newcommand{\Sbo}{Stack buffer overflow}
\newcommand{\Huaf}{Heap-use-after-free}
\newcommand{\badmalloc}{Bad-malloc}
\newcommand{\mallocdismatch}{Alloc-dealloc-mismatch}
\newcommand{\memcpyoverlap}{Memcpy-param-overlap}

\usepackage{caption}
\begin{document}

\title{Tumbling Down the Rabbit Hole: 
How do \assistfuzzingspaper{} Facilitate Grey-box Fuzzing?}
\author{\IEEEauthorblockN{Mingyuan Wu\textsuperscript{\textdagger{}\textdaggerdbl{}}}
\IEEEauthorblockA{\textit{Research Institute of Trustworthy}\\ \textit{Autonomous Systems, Southern University} \\ \textit{of Science and Technology} \\
Shenzhen, China \\
11849319@mail.sustech.edu.cn}

\and
\IEEEauthorblockN{Jiahong Xiang\textsuperscript{\textdagger{}\textdaggerdbl}}
\IEEEauthorblockA{\textit{Research Institute of Trustworthy}\\ \textit{Autonomous Systems, Southern University} \\ \textit{of Science and Technology} \\
Shenzhen, China \\
xiangjh2022@mail.sustech.edu.cn}

\and
\IEEEauthorblockN{Kunqiu Chen}
\IEEEauthorblockA{\textit{Southern University of Science and}\\ \textit{Technology} \\
Shenzhen, China \\
11911626@mail.sustech.edu.cn}

\and
\IEEEauthorblockN{Peng Di}
\IEEEauthorblockA{\textit{Ant Group}\\
Hangzhou, China \\
dipeng.dp@antgroup.com}

\and
\IEEEauthorblockN{ Shin Hwei Tan}
\IEEEauthorblockA{\textit{Concordia University}\\
Montreal, Canada \\
shinhwei.tan@concordia.ca}

\and
\IEEEauthorblockN{Heming Cui}
\IEEEauthorblockA{\textit{The University of Hong Kong}\\
Hong Kong, China \\
heming@cs.hku.hk}

\and
\IEEEauthorblockN{ Yuqun Zhang\textsuperscript{\textdagger{}}*
\thanks{\textsuperscript{*}Yuqun Zhang is the corresponding author.}
\thanks{\textsuperscript{\textdaggerdbl}These authors contributed equally.}
\thanks{\textsuperscript{\textdagger{}}These authors are also affiliated with the Department of Computer Science and Engineering, Southern University of Science and Technology, Shenzhen, China. Mingyuan Wu is also affiliated with the University of Hong Kong, Hong Kong, China.}}
\IEEEauthorblockA{\textit{Research Institute of Trustworthy}\\ \textit{Autonomous Systems, Southern University} \\ \textit{of Science and Technology} \\
Shenzhen, China \\
zhangyq@sustech.edu.cn}
}
\maketitle

\begin{abstract}
Many \assistfuzzings{} have been proposed to assist grey-box fuzzers in exploring program states guarded by tight and complex branch conditions such as equality constraints. Although they have shown promising results in their original papers, their evaluations seldom follow equivalent protocols, e.g., they are rarely evaluated on identical benchmarks. Moreover, there is a lack of sufficient investigations on the specifics of the program states explored by these strategies which can obfuscate the future application and development of such strategies. Consequently, there is a pressing need for a comprehensive study of \assistfuzzings{} on their effectiveness, versatility, and limitations to enlighten their future development. To this end, we perform the first comprehensive study about the \assistfuzzings{} for grey-box fuzzers. Specifically, we first collect \totalfuzzer{} recent fuzzers representing the mainstream \assistfuzzings{} as our studied subjects and \totalbench{} real-world projects to form our benchmark suite. After evaluating the subjects on the benchmark suite, we then surprisingly find that the \dictionarybased{} is most promising since it not only achieves similar or even slightly better performance over the other studied \assistfuzzings{} in terms of exploring program states but also is more practical to be enhanced. 
Accordingly, we propose \approach{}, which generates a customized dictionary for each seed upon the baseline fuzzer AFL to improve over the original \dictionarybased{}. The evaluation results demonstrate that \approach{} \modifyshin{increases} the edge coverage by \modifydata{16.1\%} on average for all benchmark projects over the best performer in our study (i.e., AFL++ with the \dictionarybased{}). \approach{} also successfully exposed \totalcrash{} previously unknown bugs, with \totalconfirmed{}  confirmed and \totalfixed{} fixed by the corresponding developers.
\end{abstract}

\section{Introduction}
Fuzzing has been widely adopted to expose the vulnerabilities of software systems by producing invalid, unexpected, or random data as test inputs~\cite{fuzzing}. Particularly, given a collection of seeds, grey-box fuzzers~\cite{afl, afl++, mopt, fairfuzz}  iteratively mutate them for generating new seeds to optimize their exploration on program states, i.e., executed code guarded by \branches{}~\cite{steelix}, via obtaining real-time coverage feedback based on instrumenting target programs.  


Although many grey-box fuzzers are effective in exploring sufficient program states to expose software vulnerabilities, they are still ineffective in exploring certain program states (e.g., the ones guarded by \equal{}s~\cite{firsthybrid, pbse, savior, fairfuzz}) unless developers design specific mutators for their own applications~\cite{aflcustom, libcustom}. To improve program state exploration, researchers have proposed multiple \assistfuzzings{} which are usually implemented as an independent phase in an existing fuzzer (e.g., implementing an SMT solver~\cite{Z3-SMT-solver} as a constraint-solving phase into AFL~\cite{afl}) to assist general grey-box fuzzers~\cite{havocdma,wang2019superion,redqueen,angora} for exploring such program states.  
To our best knowledge, there are \modifyshin{four} types of \assistfuzzings{}: (1) the \dictionarybased{}, which enables a list of tokens that fuzzers can insert to mutants to cope with the grammar-blind problem (i.e., generating input that may violate
grammar)~\cite{wang2019superion}, (2) the \inputstate{}, which utilizes lightweight taint tracking to monitor how input values are used at various states during program execution. In this way, the correspondences between \modifyshin{the operands} of a given instruction and the given input can be derived to first identify the offsets via lightweight taint tracking and then update their values for exploring the associated program states~\cite{redqueen, zeller}, 
(3) the \hybrid{}, which leverages SMT solvers~\cite{Z3-SMT-solver} to solve constraints that satisfy complex \branches{} for exploring program states (e.g., \qsym{}~\cite{qsym}), and (4) the \gradient{}, which utilizes gradient descent~\cite{gradient} to solve constraints for exploring program states (e.g., \angora{}~\cite{angora}).
Although these strategies have been shown effective in their corresponding papers, their evaluation can be potentially biased since they seldom follow equivalent protocols, e.g., while \angora{} and \qsym{} both use eight real-world projects in their original evaluations, they only adopt two projects (\textit{objdump} and \textit{file}) in common. 
Moreover, the evaluations of these strategies focus mainly on the performance of the grey-box fuzzers integrating the \assistfuzzings{} as a whole while neglecting the individual contributions of the strategies, e.g., the specifics of their explored program states. Without a thorough understanding of the explored program states by these strategies, it is unclear how to further improve the effectiveness of these strategies to assist grey-box fuzzers in exploring deeper program states.  

In this paper, we perform the first comprehensive study to evaluate the \assistfuzzings{}. Specifically, we first collect \totalfuzzer{} recent fuzzers as our studied subjects and construct a benchmark suite which consists of \totalbench{} open-source real-world projects commonly studied by their original papers. Then, we conduct an extensive evaluation where our evaluation results suggest that 
the \dictionarybased{} achieves quite similar and even slightly better performance over other studied \assistfuzzings{}, e.g., \afl{} activating the \dictionarybased{} slightly outperforms the best-performing SMT-solver-based fuzzer \qsym{} in terms of the average edge coverage (\modifydata{5,094} vs. \modifydata{5,067} explored edges). We observe that it is also more practical to be enhanced by strategically selecting tokens to form a dictionary for each seed. 

Inspired by our study, we propose \approach{} (\textit{\textbf{C}ustomized \textbf{D}ictionary \textbf{Fuzz}ing}), which customizes the dictionary by strategically selecting tokens for each seed upon the baseline fuzzer AFL~\cite{afl}. Specifically, \approach{} derives the execution path of each seed and extracts all its constant tokens in \equal{}s to generate a customized dictionary. Accordingly, each of such tokens is inserted to a random seed offset to generate a mutant for the further fuzzing campaign. The evaluation results indicate that under 24-hour evaluation, \approach{} can outperform the best performer in our study (i.e., AFL++~\cite{afl++} activating the \dictionarybased{}) by \modifydata{16.1\%} in terms of edge coverage. \approach{} also exposed \totalcrash{} previously unknown bugs where 30 of them can only be exposed by \approach{} \modifyshin{in our evaluation (i.e., other evaluated fuzzers can only expose 7 of them within the given time limit)}. Specifically, \totalconfirmed{} of them have been confirmed and \totalfixed{} have been fixed by the corresponding developers. 
To summarize, this paper makes the following contributions: 
\begin{itemize}
    \item To the best of our knowledge, we conduct the first comprehensive study of \totalfuzzer{} representative fuzzers on the performance impact of \assistfuzzings{} on top of a collection of real-world projects. 
    \item Our study revealed that the \dictionarybased{} is most promising because it not only achieves similar or even slightly better performance over other studied strategies, but also is more practical to be enhanced. 
    \item We propose a lightweight approach \approach{} which customizes the dictionary for each seed by strategically selecting constant tokens. It outperforms the best performer in our study by \modifydata{16.1\%} in terms of edge coverage and exposes \totalcrash{} previously unknown bugs with \totalconfirmed{} confirmed and \totalfixed{} fixed by the corresponding developers. All experimental results and our tool is publicly available in our \git{} repository~\cite{githubrepo}.
\end{itemize}

\section{Background}

\subsection{Grey-box Fuzzing}
Grey-box fuzzing~\cite{grey} has become the major practice for fuzzing. 
The objective of a grey-box fuzzer is to iteratively explore the target program thoroughly for exposing potential vulnerabilities. We take the widely-used baseline fuzzer American Fuzzy Lop (AFL)~\cite{afl} to illustrate generic grey-box fuzzers. First, AFL instruments the coverage tracking instructions into the target program in compilation time for collecting coverage information. Next, it mutates seeds (i.e., inputs of the target program) and executes the resulting mutants on the instrumented program to obtain coverage information. If such mutants explore new program states (i.e., increase code coverage), AFL identifies such mutants as ``interesting'' seeds and retains them for further exploration. The procedure above is iterated throughout the entire fuzzing campaign. While grey-box fuzzers can in general advance the exploration of program states upon the explored program states progressively, they have also been shown somewhat limited in exploring the program states guarded by tight and complex \branches{} (e.g., \equal{}s)~\cite{qsym}.  


\subsection{\assistfuzzingspaper{}}
Many \assistfuzzings{} have been proposed to assist grey-box fuzzers on exploring program states guarded by tight and complex \branches{}~\cite{qsym}~\cite{angora}~\cite{pangolin}~\cite{redqueen}. The general workflow of integrating \assistfuzzings{} in grey-box fuzzers is shown in Figure~\ref{fig:dict-framework} and illustrated as follows:  

\begin{figure}[!htb]
    \centering
    \includegraphics[width=1\columnwidth]{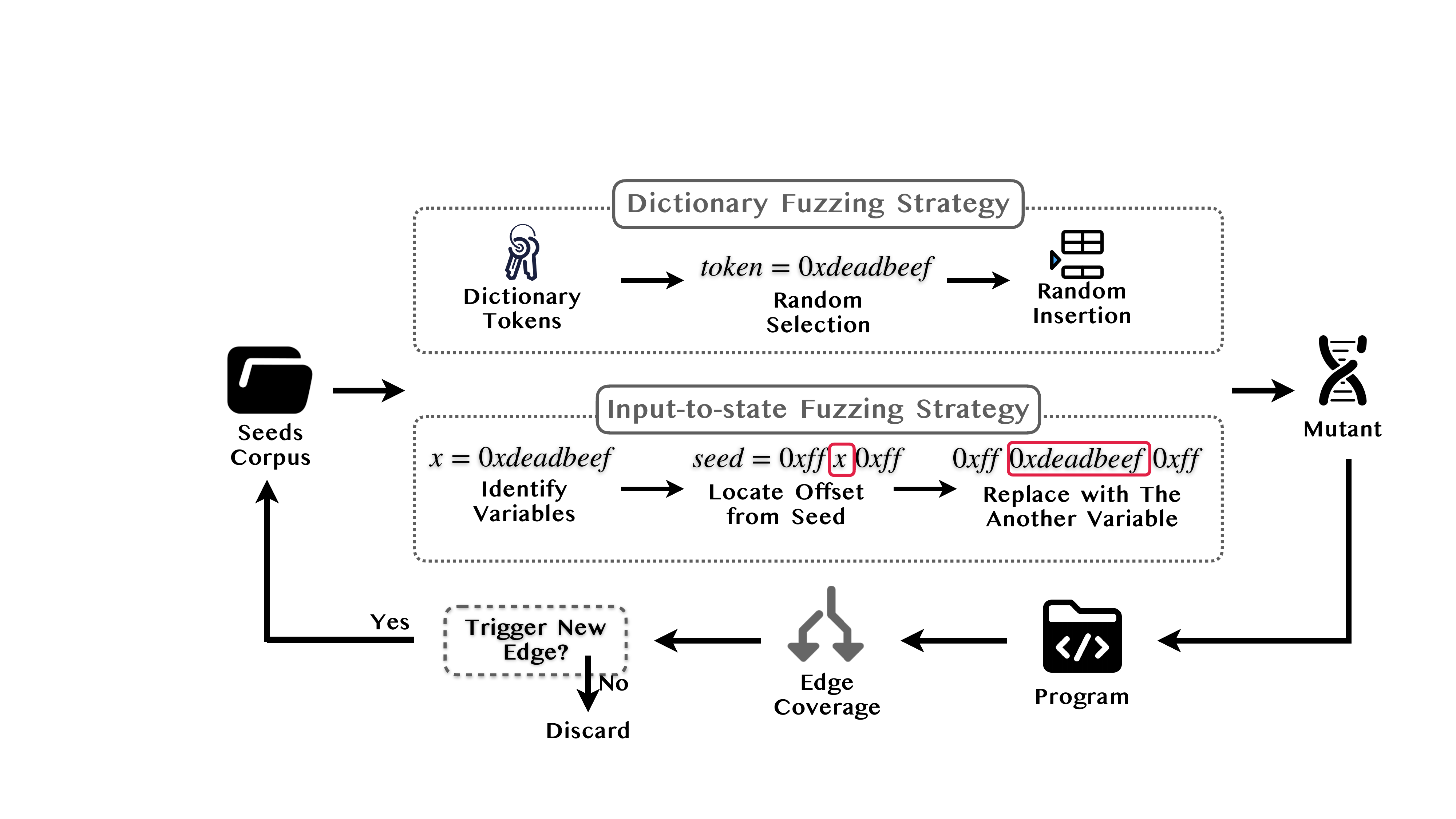}
    \caption{The workflow of \assistfuzzings{}}
    \label{fig:dict-framework}
\end{figure}

\subsubsection{The \dictionarybased{}}
\label{sec:bg:dictionary}
The \dictionarybased{} either adopts a set of user-defined tokens (so-called ``dictionary'') or generates tokens automatically by parsing the constant values defined in the target program~\cite{fuzzingdriver}.   Such tokens are randomly inserted into random seed offsets to generate mutants for the further fuzzing campaign, expecting that certain tokens can be inserted in the correct \branches{} of the target program to advance the program state exploration. 
For example in Figure~\ref{fig:dict-framework}, to explore the \equal{}  \codeIn{x == 0xdeadbeef}, the token \codeIn{0xdeadbeef} in the dictionary can be either given by users or parsed from code. Next, the token \codeIn{0xdeadbeef} is inserted to a random seed offset to generate a mutant. If the token happens to be inserted in the offset of the input corresponding to the righthand side of the \equal{} \codeIn{x == 0xdeadbeef}, i.e., the \equal{} is satisfied by executing the mutant, its guarded program state can be explored. 

\subsubsection{The \inputstate{}}\label{sec:bg:inputtostate} Since it is likely that partial input can be stored in the memory or registers at run time, the \inputstate{} derives the correspondence between such input and the associated program state to facilitate fuzzing. Specifically, the \inputstate{} identifies the lefthand and righthand side of a \branch{} during runtime via additional instrumentation. Next, it identifies which side originated from the corresponding seed, and then locates its offset via lightweight taint tracking. Furthermore, it updates such offset with the value of the other side, i.e., generating a mutant to be executed for satisfying the corresponding \branch{} such that its guarded program state can be explored.  For instance in Figure~\ref{fig:dict-framework}, given a \branch{} \codeIn{x == 0xdeadbeef}, the \inputstate{} first identifies the value of \codeIn{x}, and then traces the offset of \codeIn{x} in the seed. At last, \codeIn{x} is updated with the value \codeIn{0xdeadbeef} in the seed to generate a new mutant which is then input to the target program for exploring the program states guarded by \codeIn{x == 0xdeadbeef}. 


\subsubsection{The constraint-solving strategies (the SMT-solver-based strategy and the \gradient{})}
Symbolic execution ~\cite{klee} is widely used in grey-box fuzzers by leveraging full taint tracking and constraint solving to generate seeds satisfying target \branches{}. 
In particular, the \hybrid{} utilizes SMT-solver~\cite{Z3-SMT-solver} and the \gradient{} utilizes gradient descent~\cite{gradient} for solving constraints, respectively. 
To illustrate, in Figure~\ref{fig:dict-framework}, the grey-box fuzzer and the constraint-solving engine (i.e., the symbolic/concolic execution engine or the gradient descent solver) are first activated at the same time. Next, the grey-box fuzzer passes its generated seeds to the constraint-solving engine which leverages full taint tracking to derive constraints corresponding to the seed inputs and solve them for exploring program states guarded by tight and complex \branches{}. Meanwhile, the constraint-solving engine also passes its generated seeds to the grey-box fuzzer for the further fuzzing campaign~\cite{havocdma}. 

Although all these strategies have been well evaluated in their original papers, their evaluations can be potentially biased because (1) they seldom follow identical protocols and (2) the assessment of their individual contributions is rather obscure. Thus, there is a pressing need to comprehensively study the \assistfuzzings{} to enlighten their future development.   
\section{Empirical Study}
\subsection{Subjects \& Benchmarks}

\subsubsection{Subjects}
We aim at the grey-box fuzzers with \assistfuzzings{} as our study subjects. 
In particular, we filter many such fuzzers for selecting the representative ones. Following prior studies~\cite{havocdma, wu2022evaluating}, we first limit our search scope to the fuzzers recently published in the top Software Engineering and Security conferences (e.g., ICSE, FSE, CCS, and S\&P). Furthermore, we can only evaluate the fuzzers which are publicly available and can be successfully executed. Lastly, as it is rather challenging and time-consuming to activate certain strategies, e.g., the \dictionarybased{}, in non-AFL-based fuzzers, we only target AFL-based fuzzers. 

Finally, we select \totalfuzzer{} representative fuzzers as our studied subjects. Specifically, \afl~\cite{afl}, \aflpp~\cite{afl++}, \mopt~\cite{mopt}, and \fairfuzz~\cite{fairfuzz} represent the dictionary-based fuzzers (while they deactivate the \dictionarybased{} option by default, it can be easily activated as long as the associated tokens are provided). \qsym~\cite{qsym}, \meuzz{}~\cite{meuzz}, and \pangolin{}~\cite{pangolin} represent the SMT-solver-based fuzzers. \angora~\cite{angora} and \redqueen~\cite{redqueen} are typical gradient-based fuzzer and input-to-state-correspondence-based fuzzer, respectively.
Note that \redqueen{} is implemented into \aflpp{} (which is maintained by Google) as the \redqueen{} mode~\cite{aflpp_redqueen} and thus we choose it as the representative fuzzer of \inputstate{} following prior work~\cite{pata}.

\subsubsection{Benchmark suite}
Following prior work~\cite{havocdma}, we construct our benchmark suite based on the projects commonly adopted by the original papers of the selected fuzzers~\cite{afl++,mopt,fairfuzz,redqueen,angora,qsym}. To ensure general applicability of our study, we additionally adopt seven projects from FuzzBench~\cite{fuzzbench}, resulting in a total of 21 real-world projects.
In particular, we select 14 frequently used projects out of the papers to form our benchmark suite.
More specifically, we first select seven projects that are adopted by at least three papers. Then, we randomly select another seven projects which are adopted by one or two papers. 
The selection details are presented in our \git{} page~\cite{githubrepo}. Table \ref{tab:Benchmark-statistics} presents the statistics of our benchmark suite. Specifically, we consider our benchmark to be sufficient and representative due to the following reasons: 
\begin{enumerate}
    \item These \totalbench{} benchmark projects cover \modifydata{ten} different file formats for seed inputs, e.g., \codeIn{ELF}, \codeIn{XML}, \codeIn{JPEG}, and \codeIn{JSON}; 
    \item The sizes of these programs that range from \modifydata{1,885} to over \modifydata{150K} lines of code (LoC) can represent a wide range of programs in practice; 
    \item They are all open-source real-world programs from different vendors with various code logic. 
    \item They cover diverse functionalities including development tools (e.g., \textit{readelf}, \textit{objdump}), xml processing tools (e.g., \textit{xmlwf}), network analysis tools (e.g., \textit{tcpdump}), graphics processing tools (e.g., \textit{djpeg}), etc.
\end{enumerate}



\begin{table}[]
    \centering
    \tiny
    \caption{Statistics of the studied benchmarks}
    \label{tab:Benchmark-statistics}
    \begin{adjustbox}{width=0.95\columnwidth}
    \begin{threeparttable}
    \begin{tabular}{lllrrr}
    \hline

    \multicolumn{4}{c}{\textbf{Programs}} & \multirow{2}*{\textbf{LOC}} \\
    \cline{1-4}
    Package & Target & Commit/Version & Class \\
    \hline

                    &   readelf  &     2.40                       &   ELF  &  72,164  \\
                    &   nm       &     2.40                       &   ELF  &  55,307  \\
    binutils        &   objdump  &     2.40                       &   ELF  &  74,532  \\
                    &   size     &     2.40                       &   ELF  &  54,429  \\
                    &   strip    &     2.40                       &   ELF  &  65,432  \\ \hline
    libjpeg         &   djpeg    &      9c                        &  JPEG  &  9,023   \\ \hline
    tcpdump         &   tcpdump  &   4.99.0                       &  PCAP  &  46,892  \\ \hline
    libxml2         &   xmllint  &    2.9.12                      &   XML  &  73,320  \\ \hline
    jhead           &   jhead    &      3.04                      &  JPEG  &   1,885  \\ \hline
    libpng          &   pngfix   &     1.6.36                     &   PNG  &  12,173   \\ \hline
    libtiff         &   tiffinfo &     4.2.0                      &  TIFF  &  15,140  \\ \hline
    expat           &   xmlwf    &     2.4.8                      &   XML  &  6,871  \\ \hline
    libtiff         &   tiff2bw  &    4.2.0                       &  TIFF  &  15,024  \\ \hline
    mupdf           &   mutool   &    1.18.0                      &   PDF  & 123,575  \\ \hline
    libjpeg-turbo$^*$  &   libjpeg-turbo & 3b19db &  JPEG  &   11,106 \\ \hline
    libpng$^*$          &   libpng         & cd0ea2 &   PNG  &  31,054   \\  \hline
    libxml2$^*$         &   libxml2        & c7260a &   XML  &  104,019  \\ \hline
    re2$^*$             &   re2            & b025c6 &  REGEX &  17,754   \\ \hline
    jsoncpp$^*$         &   jsoncpp        & 8190e0 &  JSON  &    4,181   \\ \hline
    sqlite3$^*$         &   sqlite3        & c78cbf &  SQL   &   95,815   \\ \hline
    bloaty$^*$          &   bloaty         & 52948c &  ELF   &  152,845   \\ \hline
    \end{tabular} 
    \begin{tablenotes}
\item[{\tiny$*$}] These benchmark packages come from FuzzBench~\cite{fuzzbench}.
\end{tablenotes}
    \end{threeparttable}
    \end{adjustbox}
\end{table}

\subsection{Environment Setup and Implementation}
\label{sec:study:setup}
Our evaluation was conducted on the ESC servers with 128-core 2.6 GHz AMD EPYC\texttrademark{} ROME 7H12 CPUs and 256 GB RAM. The servers run on Linux 4.15.0-147-generic 64-bit Ubuntu 18.04. We strictly follow the respective original procedures of the studied fuzzers when executing them. Specifically, to allow the fuzzers to generate more tests, we set the execution time budget for all the experiments 24 hours. Meanwhile, as all fuzzers rely on randomized algorithms, we run each experiment \totalrun{} times to obtain the average result, following prior evaluations~\cite{pata, meuzz,angora}. Notably, since all the studied fuzzers are AFL-based, 
we apply the AFL (v2.57\modifyshin{b, which is the latest released version in GitHub}) llvm-mode (LLVM-13) to instrument the source code during compilation and LLVM IR~\cite{llvm} for presenting and analyzing programs. 
At last, we collect the initial seed corpus following prior work~\cite{fairfuzz, savior, pangolin, klees2018evaluating,wang2020not}.

We adopt \emph{edge coverage} to measure code coverage where an edge refers to a transition between program blocks, e.g., a conditional jump, following prior work~\cite{steelix,pangolin}. Specifically, we compute edge coverage via the unique edge number derived by the AFL built-in tool named \codeIn{afl-showmap}, which has been widely used by many existing fuzzers \cite{mopt, angora, pangolin, pata, mtfuzz,neuzz}. 
%

\noindent \textbf{Constructing \textit{Dictionary} versions via \fuzzingdriver{}.} 
To form dictionaries for all studied fuzzers involving \dictionarybased{}~\cite{static2017, fuzzingdriver}, we automatically extract tokens using the most recent \fuzzingdriver{}~\cite{fuzzingdriver} instead of relying on user-provided tokens to reduce potential bias. Specifically, \fuzzingdriver{} is designed to automatically generate dictionaries for each program, leveraging CodeQL~\cite{codeql} to extract key pieces of information from the target program's internals, including commonly occurring keywords, strings, and constants. Moreover, it employs a data cleaning module that scrutinizes extracted tokens to customize the dictionary for enhancing the efficiency of the fuzzing process. \footnote{Note that \fuzzingdriver{} is not an independent fuzzer but just a tool for extracting tokens to generate dictionaries for fuzzers, we cannot adopt it as an independent baseline in our evaluation.} 

\begin{table*}
    \caption {The average edge coverage result of studied programs}
    \label{tab:rq1:grey-box-result}
    \setlength\tabcolsep{15pt}
    \begin{adjustbox}{width=\textwidth}
    \begin{threeparttable}
    \begin{tabular}{lrrrrrrrrrrrrrr}
    \hline
    \multirow{2}*{\textbf{Benchmark}} & \multicolumn{2}{c}{\bf{AFL}} & \multicolumn{2}{c}{\bf{AFL++}}  & \multicolumn{2}{c}{\bf{\fairfuzz}} & \multicolumn{2}{c}{\bf{\mopt}} & \multirow{2}*{\textbf{\qsym}} & \multirow{2}*{\textbf{\meuzz}} & \multirow{2}*{\textbf{\pangolin}} & \multirow{2}*{\textbf{Angora}} & \multirow{2}*{\textbf{\redqueen}} \\ 
    \cline{2-9}
    &  \textbf{\original} & \textbf{\dictionary{}*}  & \textbf{\original} & \textbf{\dictionary{}*} & \textbf{\original} & \textbf{\dictionary{}*} & \textbf{\original} & \textbf{\dictionary{}*}  \\
    \hline
    
\bf{readelf}        &  10,081              &  11,561              &  10,562              &  11,524              &  11,065              &  10,954              &  11,034              &  11,757              &  11,386              &  12,136              &  11,437              &  \textbf{13,203}     &  10,053              \\   
\bf{nm}             &  4,973               &  5,199               &  5,651               &  5,270               &  4,877               &  4,830               &  4,966               &  5,580               &  \textbf{6,532}      &  5,527               &  6,394               &  5,774               &  5,526               \\   
\bf{objdump}        &  5,519               &  5,649               &  5,666               &  5,744               &  5,451               &  \textbf{6,108}      &  5,518               &  5,760               &  5,999               &  5,832               &  5,593               &  5,863               &  5,587               \\   
\bf{size}           &  3,604               &  3,800               &  4,010               &  4,015               &  3,532               &  3,551               &  3,477               &  3,618               &  \textbf{5,211}      &  4,151               &  5,045               &  5,201               &  5,002               \\   
\bf{strip}          &  6,216               &  6,369               &  6,733               &  6,598               &  6,103               &  5,714               &  5,895               &  6,210               &  \textbf{6,968}      &  6,634               &  6,741               &  5,943               &  5,513               \\   
\bf{djpeg}          &  2,319               &  2,801               &  2,542               &  2,628               &  2,446               &  2,798               &  2,304               &  2,826               &  2,092               &  2,184               &  2,238               &  \textbf{2,936}      &  2,463               \\   
\bf{tcpdump}        &  10,932              &  12,001              &  11,240              &  12,554              &  11,646              &  \textbf{13,315}     &  9,409               &  12,051              &  10,053              &  10,117              &  10,843              &  9,502               &  11,471              \\   
\bf{xmllint}        &  6,474               &  \textbf{6,862}      &  6,724               &  6,824               &  6,458               &  6,599               &  6,389               &  6,686               &  6,317               &  6,225               &  5,981               &  4,273               &  6,771               \\   
\bf{jhead}          &  156                 &  \textbf{801}        &  159                 &  772                 &  156                 &  795                 &  157                 &  728                 &  731                 &  622                 &  331                 &  157                 &  612                 \\   
\bf{pngfix}         &  1,123               &  2,237               &  975                 &  1,983               &  998                 &  2,023               &  986                 &  2,020               &  1,976               &  \textbf{2,252}      &  1,903               &  2,161               &  2,189               \\   
\bf{tiffinfo}       &  3,683               &  3,831               &  4,026               &  \textbf{4,209}      &  3,655               &  3,731               &  3,486               &  3,760               &  3,753               &  3,713               &  3,657               &  4,190               &  3,981               \\   
\bf{xmlwf}          &  5,024               &  4,990               &  4,935               &  4,953               &  4,985               &  4,993               &  4,567               &  5,025               &  4,797               &  4,498               &  5,001               &  \textbf{5,132}      &  4,879               \\   
\bf{tiff2bw}        &  3,112               &  3,503               &  3,198               &  3,478               &  3,531               &  \textbf{3,663}      &  3,102               &  3,429               &  2,871               &  3,014               &  3,128               &  3,028               &  3,615               \\   
\bf{mutool}         &  2,207               &  2,252               &  2,260               &  2,295               &  2,216               &  2,215               &  2,198               &  2,270               &  2,167               &  2,147               &  2,194               &  2,204               &  \textbf{2,314}      \\   
\bf{libjpeg-turbo}  &  4,462               &  4,596               &  4,178               &  \textbf{4,997}      &  4,900               &  4,905               &  4,997               &  4,997               &  4,722               &  4,641               &  4,785               &  4,772               &  4,554               \\   
\bf{libpng}         &  887                 &  2,092               &  884                 &  \textbf{2,263}      &  1,080               &  2,232               &  1,081               &  2,232               &  2,080               &  1,899               &  2,041               &  1,930               &  2,014               \\   
\bf{libxml2}        &  9,608               &  9,621               &  9,996               &  \textbf{11,267}     &  8,075               &  8,160               &  9,512               &  9,512               &  10,355              &  9,981               &  10,213              &  9,142               &  9,001               \\   
\bf{re2}            &  6,458               &  6,465               &  6,456               &  6,466               &  6,419               &  6,464               &  6,391               &  \textbf{6,469}      &  6,379               &  6,192               &  6,341               &  6,409               &  6,362               \\   
\bf{jsoncpp}        &  1,452               &  1,455               &  1,451               &  1,454               &  1,451               &  1,455               &  1,451               &  \textbf{1,457}      &  1,406               &  1,404               &  1,420               &  1,379               &  1,444               \\   
\bf{sqlite3}        &  6,817               &  7,989               &  6,178               &  6,286               &  6,737               &  7,925               &  7,975               &  \textbf{8,021}      &  5,013               &  4,812               &  5,143               &  3,612               &  5,588               \\   
\bf{bloaty}         &  2,043               &  2,909               &  1,912               &  3,514               &  1,912               &  2,989               &  1,980               &  3,052               &  5,595               &  \textbf{5,621}      &  5,014               &  5,410               &  5,279               \\   
\hline
\bf{Average}        &  4,626               &  5,094               &  4,749               &  \textbf{5,195}      &  4,652               &  5,020               &  4,613               &  5,117               &  5,067               &  4,933               &  5,021               &  4,868               &  4,963               \\    
\textit{p-value}        &  -              &  0.006               &  0.005               &  0.005      &  0.005               &  0.006               &  0.006               &  0.005               &  0.006               &  0.005               &  0.005               &  0.005               &  0.006               \\   

    \hline
    \end{tabular}
    \begin{tablenotes}
    \item[{}] \textbf{*Dictionaries of \textit{Fuzzer$_{Dict}$s} are all generated by FuzzingDriver~\cite{fuzzingdriver}}.
    \end{tablenotes}
    \end{threeparttable}
    \end{adjustbox}
\end{table*}

\subsection{Research Questions}
We investigate the following research questions in our study: 
\begin{itemize}
    \item \textbf{RQ1:} How well do different \assistfuzzings{} perform on our benchmark suite? 
    \item \textbf{RQ2:} What are the specifics of program states explored by  \assistfuzzings{}? 
    \item \textbf{RQ3:} What are the potential obstacles of different \assistfuzzings{}? 

\end{itemize}

\subsection{Result Analysis}

\subsubsection{RQ1: Effectiveness of the studied fuzzers}

\modifyshin{Table~\ref{tab:rq1:grey-box-result} shows the edge coverage results of all studied fuzzers where the numbers show the 
coverage results averaged over multiple runs (i.e., \totalrun{} times).} 
The ``Orig'' column denotes the original implementation of the grey-box fuzzers which deactivates the \dictionarybased{} 
and ``Dict'' denotes the grey-box fuzzers activating the \dictionarybased{} (represented as ``\textit{Fuzzer$_{Dict}$}'' in this paper). In general, we can observe that all \textit{Fuzzer$_{Dict}$}s outperform all original grey-box fuzzers in terms of the average edge coverage by \modifydata{7.9\%} to \modifydata{10.9\%}, e.g., \afldict{} explores \modifydata{10.1\%} more edges than AFL (\modifydata{5,094} vs. \modifydata{4,626} explored edges). Moreover, we also observe that \qsym, \meuzz{}, \pangolin{}, \angora{} and \redqueen outperform the best-performing grey-box fuzzer, i.e., \aflpp{}, by \modifydata{4.7\%} in terms of the average edge coverage. As all the studied subjects are AFL-based, we can derive that both \inputstate{} and \allconstraints{} can somewhat advance the exploration of program states over the original grey-box fuzzers. 
However, we can also observe that the performance advantage is limited and may not well generalize in different benchmark projects. For instance, the best-performing constraint-solving-based fuzzer \qsym{} outperforms AFL++ in \modifydata{11} projects by \modifydata{3.5\%} (\modifydata{6,968} vs. \modifydata{6,733} explored edges in \modifydata{\textit{strip}}) to \modifydata{4.6$\times$} (\modifydata{731} vs. \modifydata{159} explored edges in \modifydata{\textit{jhead}}), while \aflpp{} outperforms \qsym{} in the remaining ten projects by \modifydata{1.2\%} (\modifydata{6,456} vs. \modifydata{6,379} explored edges in \modifydata{\textit{re2}}) to \modifydata{23.2\%} (\modifydata{6,178} vs. \modifydata{5,013} explored edges in \modifydata{\textit{sqlite3}}). We also perform the Mann-Whitney U test~\cite{manntest} to demonstrate the significance of fuzzers that adopting \assistfuzzings{} compared to the grey-box fuzzer AFL. The $p$-value of less than 0.05 for AFL compared to AFL$_{Dict}$, \qsym{}, and \redqueen{} in terms of average edge coverage indicates that fuzzers adopting \assistfuzzings{} significantly outperform the grey-box fuzzer AFL.






Interestingly, we further observe from Table~\ref{tab:rq1:grey-box-result} that the dictionary-based fuzzers can potentially achieve similar or even slightly better performance over other studied fuzzers. For instance, \afldict{} outperforms \qsym{} by \modifydata{0.5\%} (\modifydata{5,094} vs. \modifydata{5,067} explored edges)  on average. Since \afldict{} and \qsym{} only differ in their adopted \assistfuzzings{}, it indicates that the \dictionarybased{} is close to or potentially more effective than the \hybrid{}. Such indications can be generalized when comparing the \dictionarybased{} with other \assistfuzzings{}. Therefore, we can infer that the dictionary-based fuzzers are effective solutions in exploring program states.

\mybox{Finding 1:The dictionary-based fuzzers achieve similar or even slightly better performance over other studied fuzzers, indicating that the \dictionarybased{} is rather effective.}

\subsubsection{RQ2: Specifics of the explored program states}
\label{rq2:analysis}
In this paper, we characterize a \emph{constraint} as a predicate that is represented as the edge between two basic blocks~\cite{basicblock} in the control-flow graph (CFG).
To reach a given \emph{program state}, all its guarded constraints should be satisfied. We thus represent each program state by an ordered sequence of its guarded constraints, i.e., a sequence of branch conditions which guard the corresponding basic blocks.
In particular, we consider the following two types of constraints based on LLVM IR: (1) an \equal{} denotes equality comparisons at the source code level (e.g., ==), which correspond to one \jumpequal{} or \jumpnoequal{}~\cite{assembly} following~\cite{equal-def}; \remove{associated with the \branch{} of an edge}(2) otherwise, it is referred to as a \nonequal{}.

\begin{figure}[!htb]
    \centering
    \includegraphics[width=0.80\columnwidth]{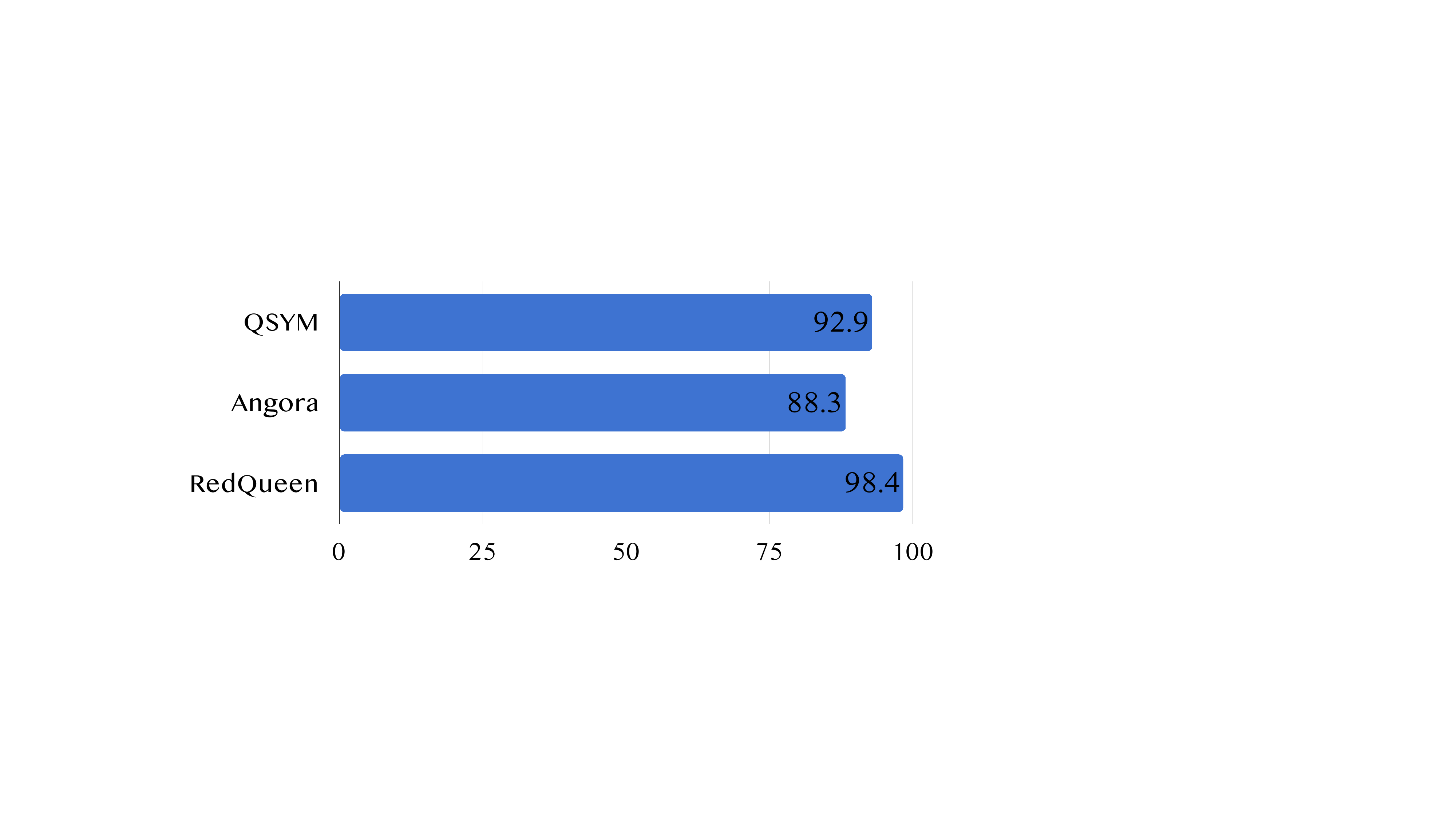}
    \caption{Ratios of the total number of \equal{}s to the total number of unsatisfiable critical constraints}
    \label{fig:rq2:find3}
\end{figure}

Our goal is to find out which constraint in the ordered sequence of a program state is critical to be satisfied by grey-box fuzzers. Particularly, given a program state represented as the ordered sequence of its guard constraints $c_1, c_2, ..., c_k, c_{k+1}, ..., c_n$ where $c_1,...,c_k$ are satisfiable and $c_{k+1}$ is unsatisfiable, we consider the first unsatisfiable constraint (i.e., $c_{k+1}$) to be \emph{critical} since it prevents the remaining unsatisfied constraints ($c_{k+1}, ..., c_n$) from being explored. 
We first investigate the specifics of the program states which can be explored by the fuzzers with \assistfuzzings{} other than the grey-box fuzzers to reflect the improvement that the \assistfuzzings{} brings to grey-box fuzzers. Specifically, we collect the unexplored program states (denoted as $S$) by applying the grey-box fuzzers. We then filter out the ones which cannot be explored by any fuzzer with \assistfuzzings{}. Finally, we derive the \emph{critical constraints} from the remaining program states for further analysis. We observe that the majority of the constraints which can be explored by the fuzzers integrating \assistfuzzings{} 
but cannot be explored by their corresponding grey-box fuzzers are \equal{}s. 
\modifycr{\remove{Figure~\ref{fig:rq2:find3} presents the ratios of the total number of \equal{}s to the total number of critical constraints. For example, \modifydata{98.1\%} critical constraints explored by  AFL$_{Dict}$ other than AFL are \equal{}s, \modifydata{97.4\%} by AFL++$_{Dict}$ other than AFL++, 
and \modifydata{93.1\%} by \qsym{} other than AFL.}} Figure~\ref{fig:rq2:find3} presents the ratios of the total number of \equal{}s to the total number of critical constraints. For example, 98.1\% of the critical constraints explored by AFL$_{Dict}$, 97.4\% by AFL++$_{Dict}$, and 93.1\% by \qsym{}, but not by their corresponding grey-box fuzzers, are equality constraints.

\mybox{Finding 2: The \assistfuzzings{} bring improvement over grey-box fuzzers by effectively exploring program states guarded by \equal{}s.}


We further investigate the characteristics for the unexplored program states by \textit{Fuzzer$_{Orig}$}s. Note that all these program states 
are guarded by \equal{}s and can be explored by either \dictionarybased{}, \inputstate{}, or \allconstraints{}. Specifically, we use a semi-automated approach to analyze the program state characteristics guarded by \equal{}s: (1) our script first automatically extracts \equal{}s, and then (2) we manually identify common characteristics among these \equal{}s. Surprisingly, we find that \modifydata{92.3\%} of these unexplored \equal{}s share the same form as \codeIn{input [==|!=]} CONSTANT (e.g., function \codeIn{memcmp(input, CONSTANT)} or \codeIn{switch(input)\{case CONSTANT\}}) after compilation, i.e., taking the content directly from the input to compare with a predefined constant value (constant-evaluating \equal{}s). 
\mybox{Finding 3: The constraints explored by \dictionarybased{} and other studied strategies are mostly constant-evaluating \equal{}s.}

\subsubsection{RQ3: The obstacles of different strategies}
\label{sec:rq3:analysis}
Our previous findings indicate that all \assistfuzzings{}  achieve similar edge coverage performance. 
We then discuss the potential obstacles for their future development to understand which strategy is more practical to be enhanced for better exploring the program states guarded by tight and complex \branches{}. 

\begin{figure}[!htb]
    \centering
    \includegraphics[width=0.95\columnwidth]{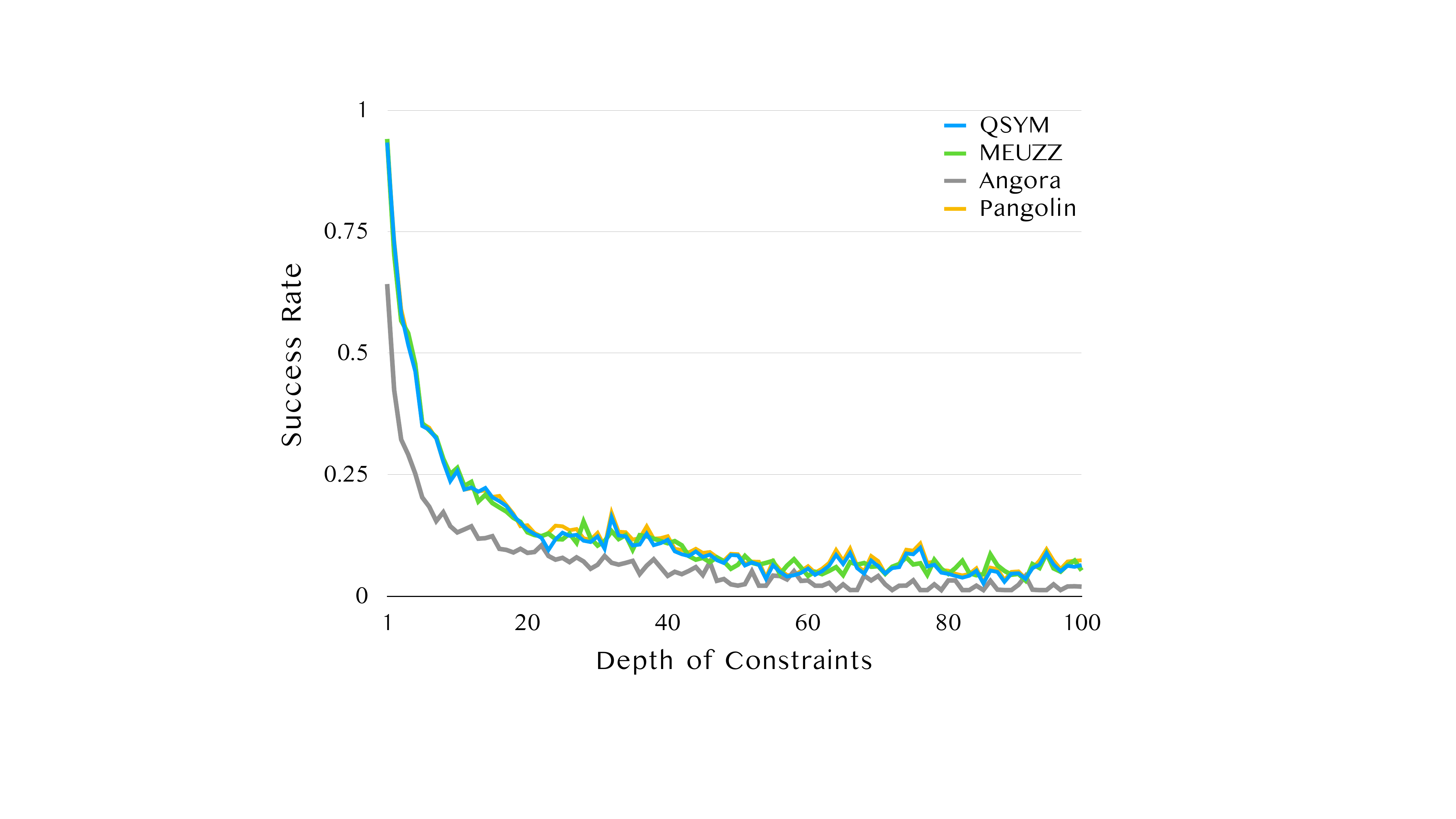}
    \caption{Ratios of the execution time and explored edges of the \inputstate{}}
    \label{fig:requeen_ratio}
\end{figure}
\noindent
\textbf{\Inputstate{}.} For the \inputstate{}, we first investigate its potential effect by studying the input-to-state-correspondence-based fuzzer, i.e., \redqueen{}. Apart from \inputstate{}, \redqueen{} also includes other strategies (e.g., the \textit{havoc} strategy). We then investigate the ratio of the execution time of the \inputstate{} to the overall execution time budget (i.e., 24 hours) for studying its effectiveness, as in Figure~\ref{fig:requeen_ratio}.  
We can observe considerable variations on the execution time of the \inputstate{} across different benchmarks. For instance, in \textit{readelf}, \textit{nm}, and \textit{strip}, its execution exceeds 20 hours. On the contrary, in \textit{jhead} and \textit{xmlwf}, its execution lasts  merely less than one hour. Meanwhile, Figure~\ref{fig:requeen_ratio} also presents the ratio of the edges explored by the \inputstate{} to the overall explored edges. It is surprising to see that executing the \inputstate{} longer does not necessarily explore more edges, e.g., executing 21 hours but only exploring 21.5\% edges in \textit{readelf}, 
indicating that the adopted mechanisms (e.g., lightweight taint analysis) in the \inputstate{} can incur performance issues in certain benchmark projects. We further infer that its development may be hindered by the inaccuracies of its adopted lightweight taint tracking. To illustrate, we manually analyze \modifydata{10\% (271) of} constant-evaluating \equal{}s which \redqueen{} fails to explore in our evaluation. We observe that although it successfully detected the corresponding constant values for these \equal{}s, its lightweight taint analysis failed to locate their corresponding offsets in the seeds. Figure~\ref{fig:xmllint-cmp-example} presents one such example from \textit{xmllint} where \redqueen{} successfully obtains the constant value but it fails to locate its offset corresponding to \codeIn{CUR\_PTR} in the seed. Therefore, the \equal{} at line 3 cannot be satisfied, and \redqueen{} fails to explore its guarded program states. 

\mybox{Finding 4: The \inputstate{} could potentially trigger performance issues in certain benchmarks while also running the risk of inaccuracies in locating correct offsets.}

\begin{figure}[htb]
    \lstset{style=demo-code}
    \centering
    \begin{lstlisting}[language=C, basicstyle={\fontsize{6.5}{6.5}\ttfamily}]
void xmlParseNotationDecl(xmlParserCtxtPtr ctxt) {
  // CMP10 checks if the first 10 characters of a string 's' match the 10 provided characters (c1 to c10).
  |\uncovered|if(CMP10(CUR_PTR,'<','!','N','O','T','A','T','I','O','N')) {
    // ... 
}} 
\end{lstlisting}
    \caption{An input-to-state fuzzing strategy failure case in \textit{xmllint parser.c}}
    \label{fig:xmllint-cmp-example}
\end{figure}

\begin{figure}[htb]
    \lstset{style=demo-code}
    \centering
    \begin{lstlisting}[language=C, basicstyle={\fontsize{6.5}{6.5}\ttfamily}]
if (alias != XML_CHAR_ENCODING_ERROR) {
    const char* canon;
    // The return value of xmlGetCharEncodingName is 
    // determined by a switch statement
    canon = xmlGetCharEncodingName(alias);
    |\colorbox{uncoveredgray}{if ((canon != NULL) \&\& (strcmp(name, canon)))}|{
       return(xmlFindCharEncodingHandler(canon)); 
}}
    \end{lstlisting}
    \caption{A case in \textit{xmllint encoding.c}}
    \label{fig:hybrid-disable}
\end{figure}



\begin{figure}[!htb]
    \centering
    \includegraphics[width=0.82\columnwidth]{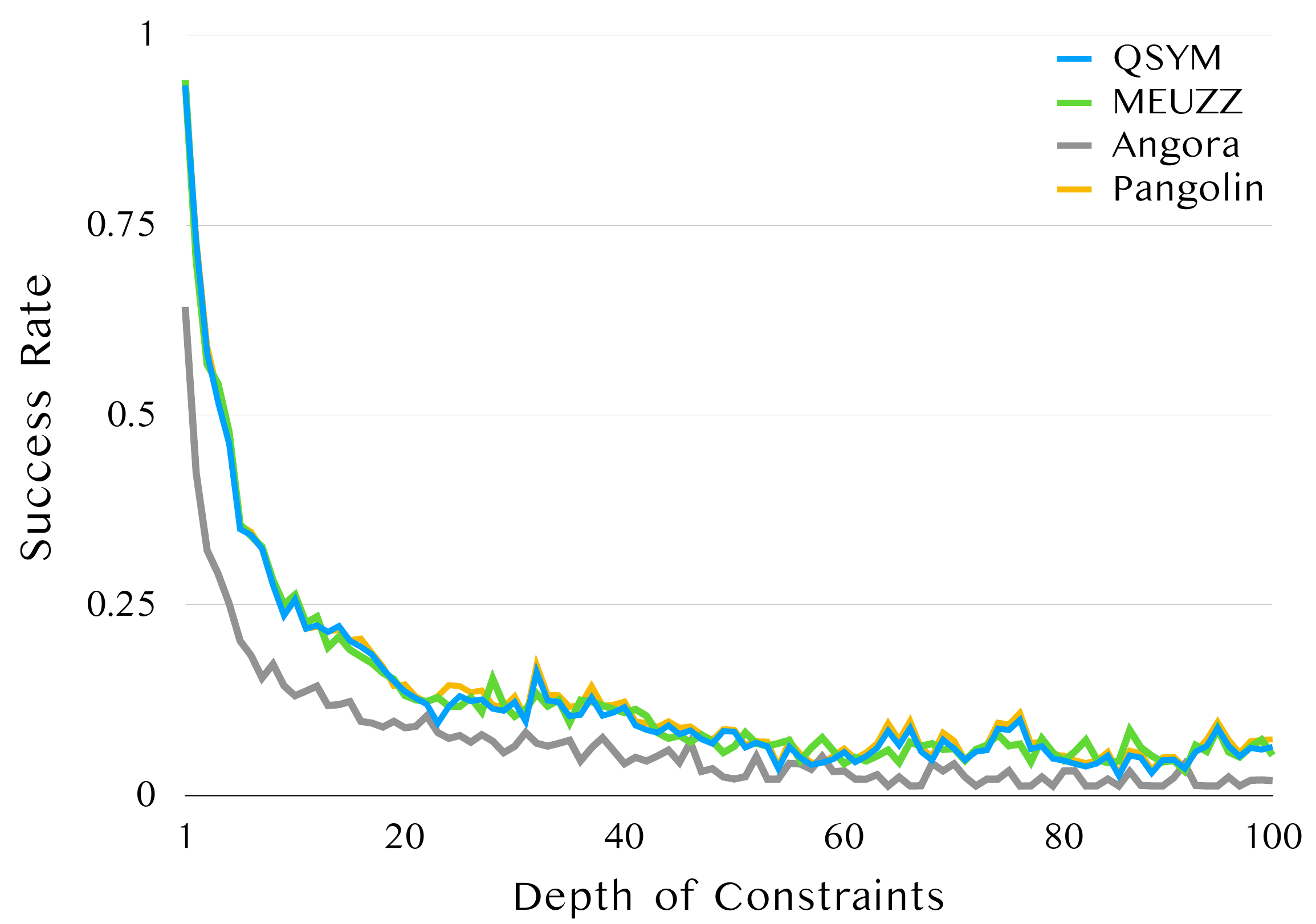}

    \caption{The success rates of solving constraints at different depths}
    \label{fig:sat-ratio}
\end{figure}

\noindent
\textbf{\Allconstraints{}}.
For the constraint-solver-based strategies (i.e., the \hybrid{} and the \gradient{}), we infer that although they can advance the exploration of program states to some extent, their effectiveness can nevertheless be compromised when exploring program states guarded by certain complex and tight \branches{}. Notably, the constraint-solver-based fuzzers fail to explore many program states guarded by \equal{}s which can otherwise be explored by \textit{Fuzzer$_{Dict}$}s. For example in Figure~\ref{fig:hybrid-disable}, only the \textit{Fuzzer$_{Dict}$}s can explore the condition in line 6 \codeIn{(canon != NULL) \&\& ...)} in our study. 
To investigate how the performance of the constraint-solver-based strategies correlates with the specifics of program states, we first characterize the ``depth'' of constraints. As the constraint $c_{k+1}$ can be reached if and only if the conjunction of its all dependent constraints $c_1 \land c_2 \land ...\land c_k$ is satisfied, we thus compute the ``depth'' for constraints $c_{k+1}$ as the size $k$ of the conjunction of its dependent constraints. Figure~\ref{fig:sat-ratio} demonstrates the success rate of solving constraints at different depths by applying the \allconstraints{}, where ``successful solving'' refers to that a constraint-solver finds a satisfiable solution for these constraints. We observe that for all constraint-solver-based fuzzers, the success rate significantly decreases as the depth approaches 20, e.g., \qsym{} has 23.6\% success rate when the depth reaches 10 while only 15.0\% when reaching 20. 

\mybox{Finding 5: Although constraint-solver-based strategies may potentially explore tight and complex constraints, it still becomes less effective when solving deeper constraints.}

\begin{figure}[!htb]
    \centering
    \includegraphics[width=0.75\columnwidth]{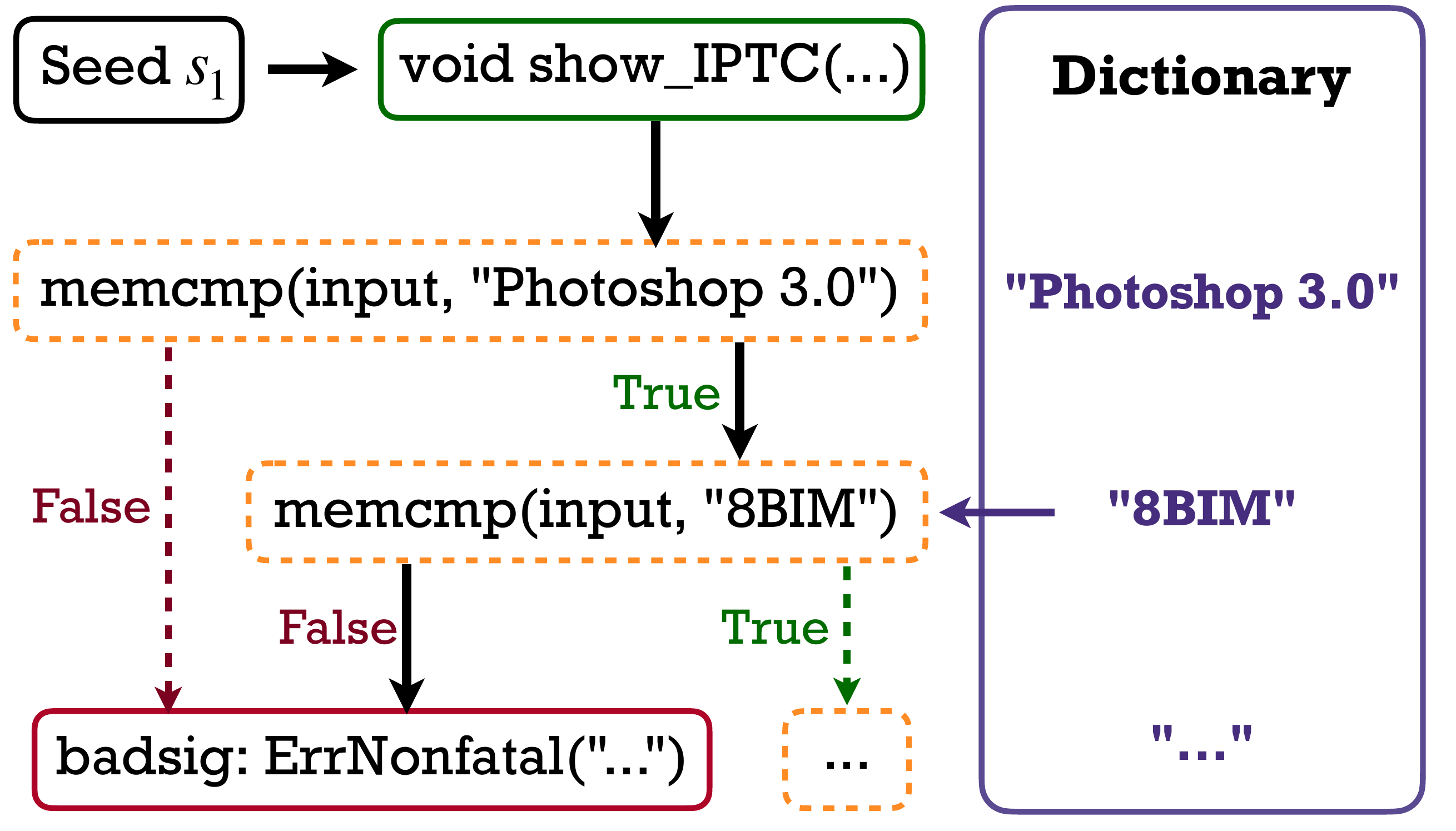}
    \caption{Strategically building a dictionary for exploring \textit{jhead iptc.c}}
    \label{fig:sibling-tokens}
\end{figure}

\noindent
 \textbf{\Dictionarybased{}.}
We then investigate the potential obstacles which may hinder the future development of the \dictionarybased{}. Interestingly, we observe that although the \textit{Fuzzer$_{Dict}$}s can slightly outperform the fuzzers with other \assistfuzzings{} in terms of the average edge coverage, they fail to achieve consistent performance advantages in each benchmark project. For instance,  \qsym{} and \angora{} outperform the best performer of \textit{Fuzzer$_{Dict}$}s, i.e., AFL++$_{Dict}$, in \modifydata{nine} benchmark projects by \modifydata{2.1\%} to \modifydata{59.2\%}. We thus infer that the power of the \dictionarybased{} has not been fully exploited, i.e., randomly selecting tokens to form a dictionary may be essentially deficient. To validate this hypothesis, we set out to evaluate what performance impact can be caused by strategically selecting the tokens to form a dictionary. Specifically, we first randomly selected \modifydata{10\% (268)} of \equal{}s out of our benchmark suite which can be explored by the fuzzers with all the \assistfuzzings{} other than the \dictionarybased{}. Next, for an edge in CFG corresponding to the failed exploration on an \equal{} (e.g., the \codeIn{False} edge from \codeIn{memcmp(input, "8BIM")} in Figure~\ref{fig:sibling-tokens}) when running a seed (e.g., $s_1$ in Figure~\ref{fig:sibling-tokens}), we identify its ``sibling'' edges (i.e., edges under one shared prefix edge defined in \cite{neuzz}, such as the \codeIn{True} edge from \codeIn{memcmp(input, "8BIM")} in Figure~\ref{fig:sibling-tokens}). Accordingly, we only collect the associated constant tokens (e.g., \codeIn{"8BIM"}) 
and add them into the dictionary of the seed such that it can be randomly inserted to its offsets. At last, we apply \aflpp{}$_{Dict}$ to run the target program. As a result, \modifydata{93.7\% (251)} of such \equal{}s can be explored by \aflpp{}$_{Dict}$ in three hours, indicating that strategically selecting tokens to form a dictionary can advance the exploration of the program states guarded by \equal{}s.  

\mybox{Finding 6: The exploration of program states guarded by the \equal{}s can be enhanced by strategically choosing tokens to form a customized dictionary.}

\subsection{Discussion}
As \assistfuzzings{} have been shown to be powerful and yet limited in our study, we then discuss how we could potentially enhance \assistfuzzings{} for advancing program state exploration in practice, which essentially demands being lightweight.
While it has been widely recognized that improving the taint analysis and constraint solvers can be typically heavyweight to cause potentially excessive efforts~\cite{redqueen,profuzzer,greyone}, strategically selecting tokens to form a dictionary for the \dictionarybased{} is likely to be lightweight and more practical. Note that while ideally, accurately tracking the offsets of seeds to insert the correct tokens can further improve the effectiveness of the \dictionarybased{}, it is nevertheless heavyweight as presented in prior work~\cite{redqueen,profuzzer,greyone}. We thus do not consider accurately tracking the seed offsets to improve over the \dictionarybased{}. 


\section{Customized Dictionary Fuzzing}

Motivated by our previous findings, we propose \approach{} (\textit{\textbf{C}ustomized \textbf{D}ictionary \textbf{Fuzz}ing}) which builds upon the baseline fuzzer AFL a customized dictionary for each seed by accurately selecting tokens.

\subsection{Approach}
 
\begin{figure}[!htb]
    \centering
    \includegraphics[width= 0.95 \columnwidth]{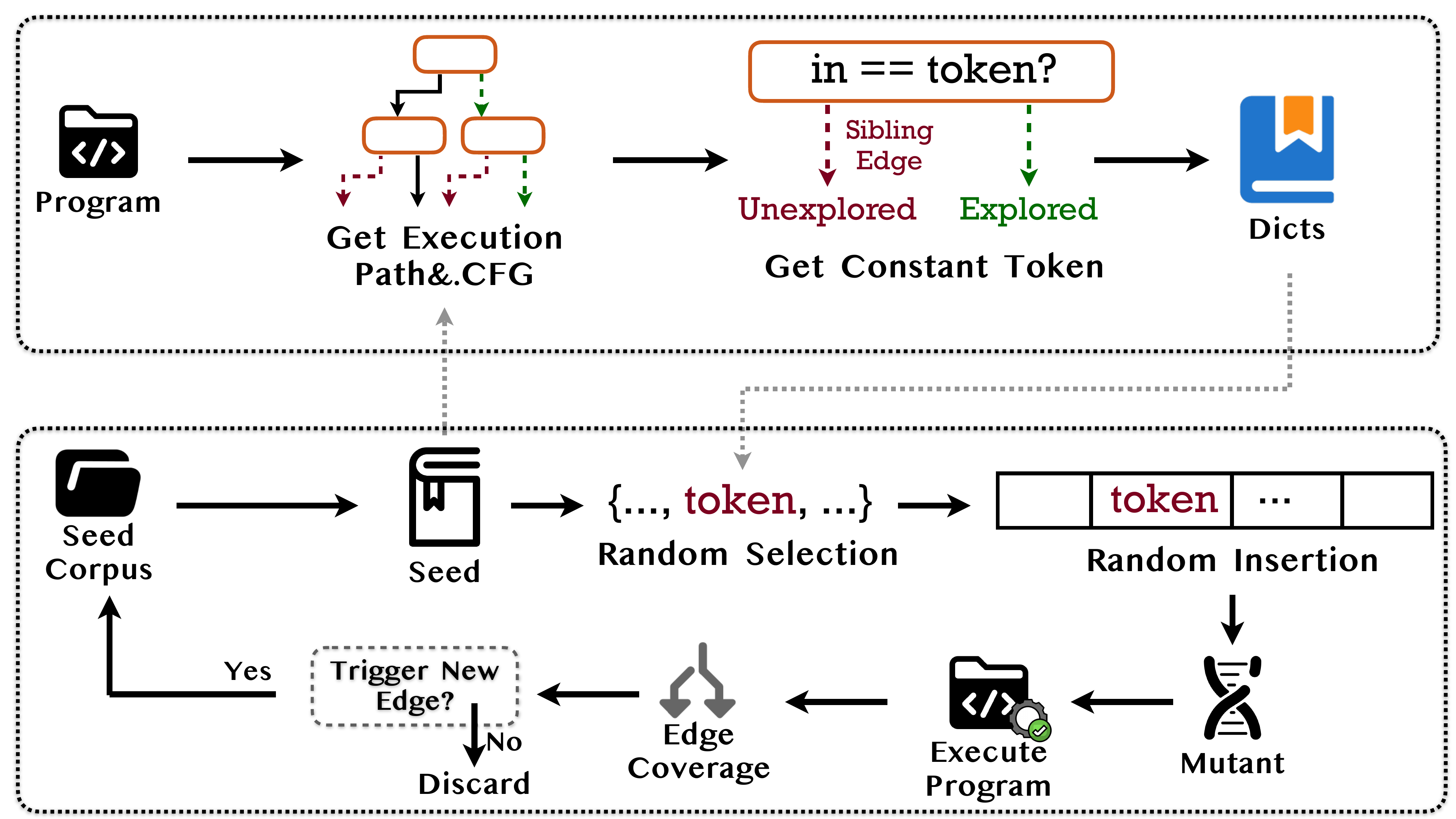}
    \caption{The workflow of \approach{}}
    \label{fig:dict-workflow}
\end{figure}

 Figure~\ref{fig:dict-workflow} presents the workflow of \approach{}. For each seed in the seed corpus, \approach{} first derives its execution path, and then extracts all its constant tokens in \equal{}s to form a dictionary for the seed. 
 Next, \approach{} randomly selects a token from the dictionary and inserts it into a random offset to generate a mutant. If running such a mutant upon the target program increases edge coverage, it will be added to the seed corpus.

\begin{algorithm}[!t]
  \caption{Customized Dictionary Fuzzing}
  \label{alg:framework}
\begin{flushleft}
  \hspace*{\algorithmicindent}  \textbf{Input}:  \textit{initialSeed}, \textit{budget}
  \end{flushleft}
  \begin{flushleft}
  \hspace*{\algorithmicindent} \textbf{Output}: \textit{seedCorpus}
  \end{flushleft}
  \small{
	\begin{algorithmic}[1]
    \Function{CustomizedDictionaryFuzzing}{}
    \State \textit{CFG} $\leftarrow$ \textit{getCFGFromTargetProgram}()
    \State \textit{tokens} $\leftarrow$ \textit{getConstantTokensFromCFG}(\textit{CFG})
    \State \textit{path} $\leftarrow$ \textit{getPathBySeed}(\textit{initialSeed}, \textit{CFG})
    \State \textit{dict} $\leftarrow$ \textit{getValidToken}(\textit{tokens}, \textit{path}, \textit{CFG})
    \State \textit{dicts} $\leftarrow$ $[\![$
    \textit{initialSeed} $\Rightarrow$ \{\textit{dict}\}$]\!]$
    \State \textit{seedCorpus} $\leftarrow$ \{\textit{initialSeed}\}
    \While{\textit{fuzzing time} \textbf{not} \textit{exceed budget}} 
    \For{\textit{each seed} \textbf{in} \textit{seedCorpus}}
    \State \textit{sDict} $\leftarrow$ \textit{dicts}$[\![$\textit{seed}$]\!]$ 
    \State \textit{token} $\leftarrow$ \textit{randomSelection}(\textit{sDict}) 
    \State \textit{mutant} $\leftarrow$ \textit{randomInsertion(seed, token)}
    \If{\textit{mutant has new edges}}
    \State \textit{seedCorpus} $\leftarrow$ \textit{seedCorpus} $\cup$ \{\textit{mutant}\}
    \State \textit{muPath} $\leftarrow$ \textit{getPathBySeed}(\textit{mutant}, \textit{CFG})
    \State \textit{muDict} $\leftarrow$ \textit{getValidToken}(\textit{tokens}, \textit{muPath}, \textit{CFG})
    \State \textit{dicts}$[\![$\textit{mutant}$]\!]$ $\leftarrow$ \textit{muDict}
    \EndIf
    \EndFor
    \EndWhile
    \State \textbf{return} seedCorpus
    \EndFunction
	\end{algorithmic}
    }
\end{algorithm}
\modifyshin{Instead of maintaining an overall dictionary as in prior dictionary-based approaches~\cite{afl,afl++,mopt,fairfuzz} , \approach{} generates a customized dictionary for each seed input. Our intuition is that having a separate dictionary for each seed will (1) allow easier tracking and effective selection of relevant tokens for each seed, and (2) avoid polluting or overloading the dictionary with tokens from other seeds.} Algorithm~\ref{alg:framework} presents the workflow of \approach{}, which takes an \codeIn{initialSeed} as input, and performs fuzzing with a given time \codeIn{budget}. First, we parse the control-flow graph of the target program \codeIn{CFG} and its corresponding constant tokens following previous work~\cite{fuzzingdriver} (lines 2--3). Next, we initialize the seed corpus \codeIn{seedCorpus}  by parsing all the constant tokens extracted from the \equal{}s of the executed path \codeIn{path}
of \codeIn{initialSeed} via the \codeIn{getValidToken} function. Specifically, this function extracts constant tokens from the ``sibling'' edges out of all the edges corresponding to failed exploration on \equal{}s of \codeIn{path}. In this way, we form a customized dictionary for \codeIn{initialSeed} with the collected tokens. 
For each \codeIn{seed}, \approach{} generates a mutant for exploring new program states. To avoid inserting irrelevant tokens into the \codeIn{seedCorpus}, we only select tokens from the customized dictionary of such \codeIn{seed} to generate a \codeIn{mutant} (lines 10--12). Meanwhile, if running such a \codeIn{mutant} successfully explores new program states, we add it to \codeIn{seedCorpus} for further exploration (lines 13--14). Similarly, we generate a customized dictionary for this \codeIn{mutant} by parsing the constant tokens according to its executed path (lines 15--17). For example, assuming that running a seed presented in Figure~\ref{fig:sibling-tokens} fails to satisfy the \equal{} \codeIn{memcmp(input, "8BIM")}, \approach{} then identifies the associated constraint-solving constant token \codeIn{"8BIM"} while filtering out other irrelevant tokens (e.g., \codeIn{"Photoshop 3.0"}) to customize a dictionary for further mutations. 

\subsection{Evaluation}
To evaluate \approach{}, we include the best-performing dictionary-based fuzzer AFL++$_{Dict}$ and constraint-solver-based fuzzer \qsym{} as well as the input-to-state correspondence fuzzer \redqueen{} for performance comparison. We also include AFL$_{Dict}$ to assess the effectiveness of the customized dictionary by \approach{} since they only differ in the adopted dictionaries. 
Similar to the setup in Section~\ref{sec:study:setup}, we run each experiment \totalrun{} times to obtain the average result within 24 hours. Note that the dictionary for each seed is built on-the-fly so the time cost is included in the overall running time (24 hours).
Prior to running Algorithm 1, \approach{} first builds CFG and collects constant tokens (this process only incurs roughly 10 seconds overhead per benchmark which is minimal compared to \codeIn{afl-clang-fast}).  We further present the details of compilation cost in our \git{} page~\cite{githubrepo} due to the page limit.

\begin{table}
    \caption {The edge coverage results of \approach}
    \label{tab:approach-coverage-result}
    \setlength\tabcolsep{15pt}
    \begin{adjustbox}{width=\columnwidth}
    \begin{tabular}{lrrrrrrrr}
    \hline
    \textbf{Benchmark}   & \bf{AFL$_{Dict}$} & \textbf{AFL++$_{Dict}$} & \bf{\qsym} & \bf{\redqueen} & \textbf{\approach} \\ 
    \cline{2-4}
    \hline
    
\bf{readelf}        &  11,561              &  11,524              &  11,386              &  10,053              &  \textbf{13,276}     \\   
\bf{nm}             &  5,199               &  5,270               &  6,532               &  5,526               &  \textbf{6,719}      \\   
\bf{objdump}        &  5,649               &  5,744               &  5,999               &  5,587               &  \textbf{6,539}      \\   
\bf{size}           &  3,800               &  4,015               &  \textbf{5,211}      &  5,002               &  5,048               \\   
\bf{strip}          &  6,369               &  6,598               &  6,968               &  5,513               &  \textbf{8,327}      \\   
\bf{djpeg}          &  2,801               &  2,628               &  2,092               &  2,463               &  \textbf{2,874}      \\   
\bf{tcpdump}        &  12,001              &  12,554              &  10,053              &  11,471              &  \textbf{14,744}     \\   
\bf{xmllint}        &  6,862               &  6,824               &  6,317               &  6,771               &  \textbf{7,830}      \\   
\bf{jhead}          &  801                 &  772                 &  731                 &  612                 &  \textbf{865}        \\   
\bf{pngfix}         &  2,237               &  1,983               &  1,976               &  2,189               &  \textbf{2,342}      \\   
\bf{tiffinfo}       &  3,831               &  4,209               &  3,753               &  3,981               &  \textbf{4,521}      \\   
\bf{xmlwf}          &  \textbf{4,990}      &  4,953               &  4,797               &  4,879               &  4,980               \\   
\bf{tiff2bw}        &  3,503               &  3,478               &  2,871               &  3,615               &  \textbf{4,728}      \\   
\bf{mutool}         &  2,252               &  2,295               &  2,167               &  2,314               &  \textbf{2,333}      \\   
\bf{libjpeg-turbo}  &  4,596               &  4,997               &  4,722               &  4,554               &  \textbf{5,001}      \\   
\bf{libpng}         &  2,092               &  2,263               &  2,080               &  2,014               &  \textbf{2,264}      \\   
\bf{libxml2}        &  9,621               &  11,267              &  10,355              &  9,001               &  \textbf{12,398}     \\   
\bf{re2}            &  6,465               &  6,466      &  6,379               &  6,362               &  \textbf{6,470}               \\   
\bf{jsoncpp}        &  1,455               &  1,454               &  1,406               &  1,444               &  \textbf{1,459}      \\   
\bf{sqlite3}        &  7,989               &  6,286               &  5,013               &  5,588               &  \textbf{8,209}      \\   
\bf{bloaty}   &  2,909               &  3,514               &  5,595               &  5,279               &  \textbf{5,771}      \\   
\hline
\bf{Average}        &  5,094               &  5,195               &  5,067               &  4,963               &  \textbf{6,033}      \\ 
    \textit{p-value}     &   0.005                 &  0.005            &  0.005            &  0.005                   &         -            \\ 

    \hline
    \end{tabular}
    \end{adjustbox}
\end{table}
\subsubsection{Result and analysis}
Table~\ref{tab:approach-coverage-result} presents the edge coverage results of the studied approaches on top of all the benchmark projects. In general,  by only differing the adopted dictionaries, \approach{}  significantly outperforms AFL$_{Dict}$ by \modifydata{18.4\%} (\modifydata{6,033} vs. \modifydata{5,094} explored edges), indicating the effectiveness of our proposed customized dictionary. Moreover, \approach{} outperforms AFL++$_{Dict}$ by \modifydata{16.1\%} (\modifydata{6,033 vs. 5,195} explored edges), \qsym{} by \modifydata{19.1\%}  (\modifydata{6,033 vs. 5,067} explored edges), and \redqueen{} by \modifydata{21.6\%} (\modifydata{6,033 vs. 4,963} explored edges) averagely. The results suggest that \approach{} can significantly improve the effectiveness of the \dictionarybased{}. We also perform the Mann-Whitney U test~\cite{manntest} to illustrate the significance of \approach{}. The fact that the $p$-value of \approach{} comparing with AFL$_{Dict}$ in terms of the average edge coverage is 0.00503 indicates that \approach{} outperforms AFL$_{Dict}$ significantly ($p$ $<$ 0.05). 

\begin{figure}[!htb]
    \centering
    \includegraphics[width=0.85\columnwidth]{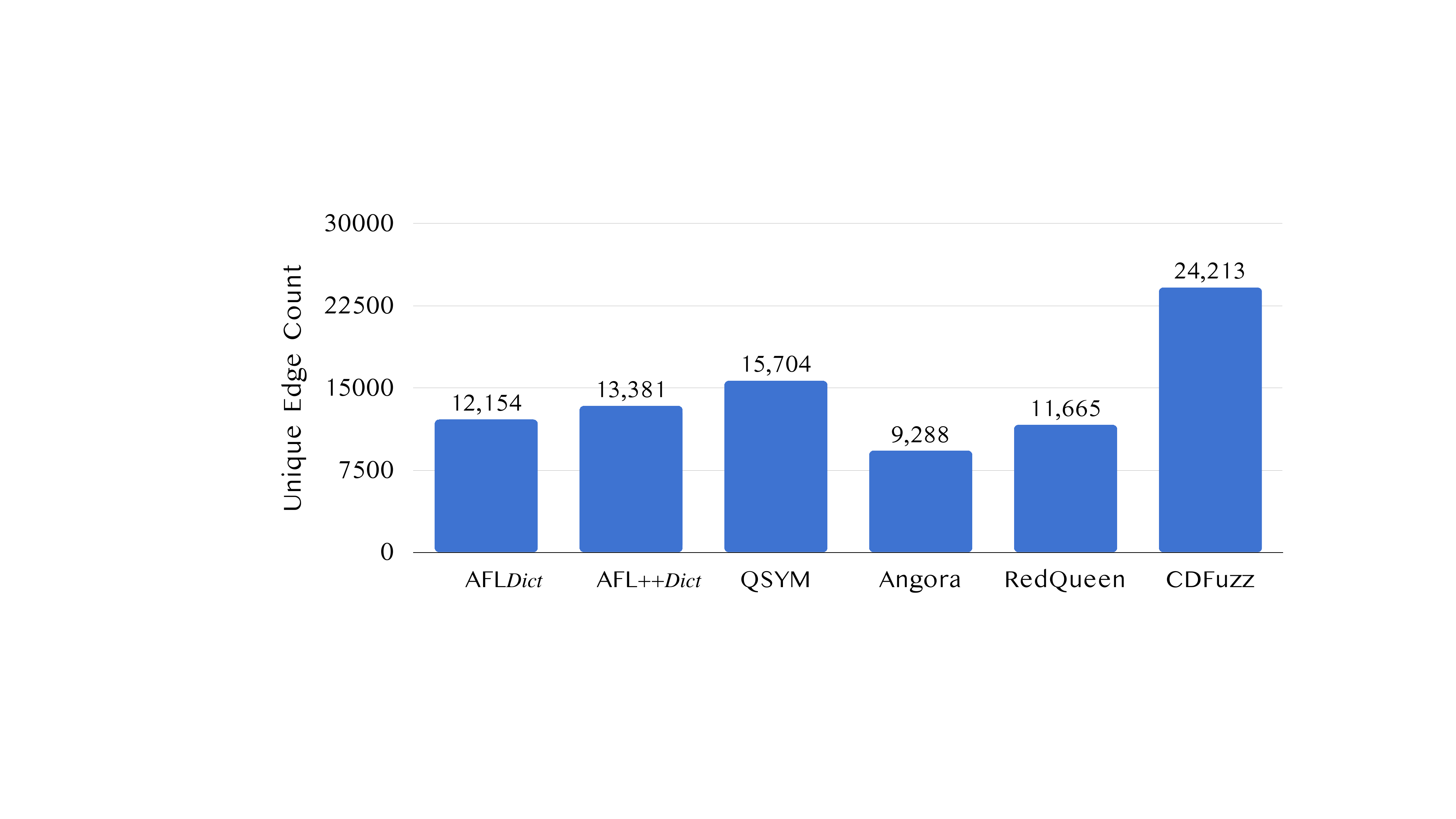}
    \caption{The explored \equal{}s of each studied fuzzer}
    \label{fig:unique-edge-cnt}
\end{figure}

\begin{figure*}[h] 
\includegraphics[width=0.95 \textwidth]{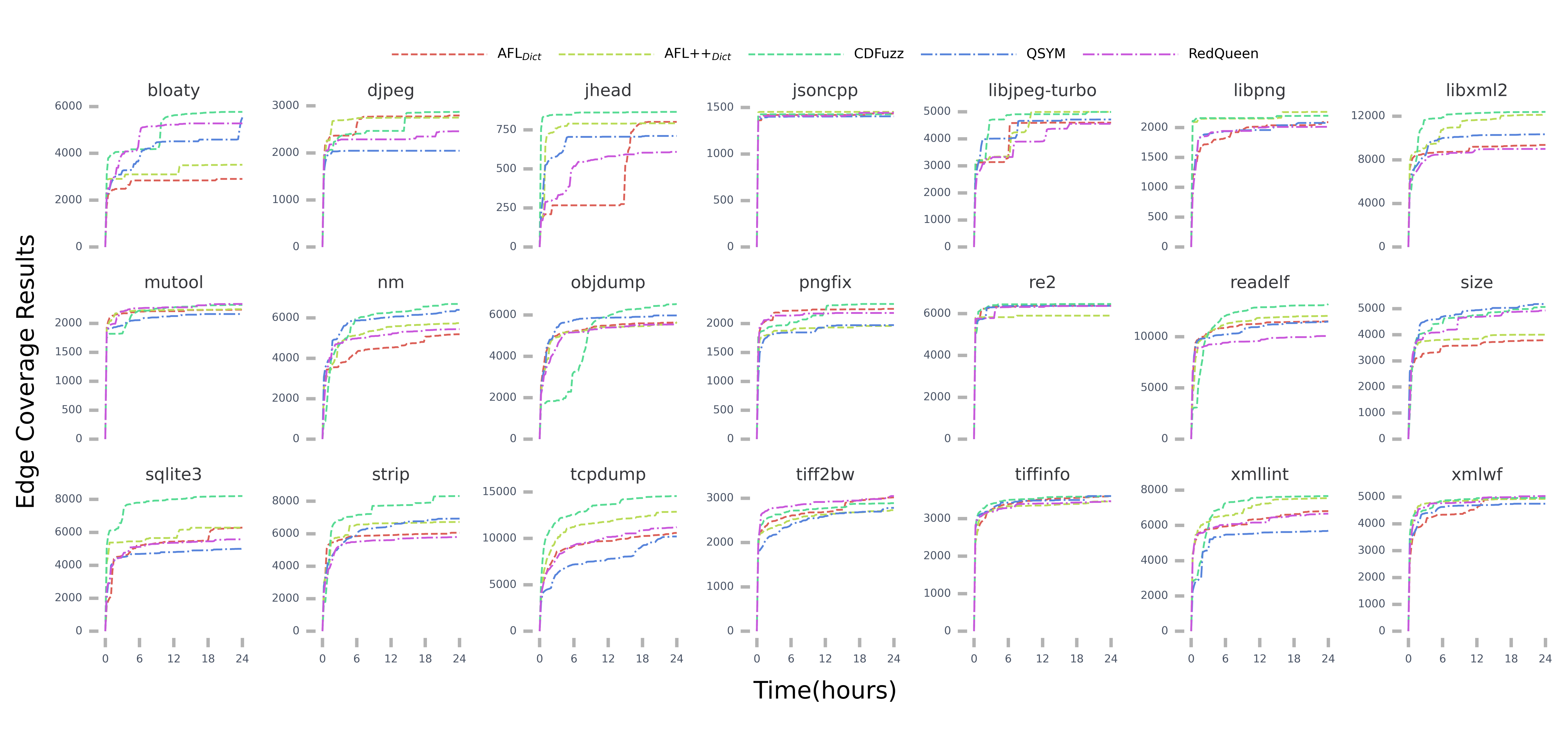}
\centering
\caption{Edge coverage of \approach{} over time}
\label{fig:dict-trend}
\end{figure*}

We further investigate how the fuzzers perform on exploring \equal{}s, as in Figure~\ref{fig:unique-edge-cnt}. Specifically, \approach{} outperforms the runner-up performer \qsym{} by \modifydata{29.3\%} in terms of exploring \equal{}s (\modifydata{16,145 vs. 12,484}). The results indicate that \approach{} can significantly improve the power of exploring \equal{}s for grey-box fuzzing.

Figure~\ref{fig:dict-trend} presents the edge coverage trends of the studied approaches in each benchmark within 24 hours. 
Overall, \approach{} outperforms all studied fuzzers significantly in most of the benchmarks (except \modifydata{\textit{xmlwf}, and \textit{size}}). Specifically, \approach{} outperforms the best-performing AFL++$_{Dict}$ of our study in all benchmark projects  in terms of edge coverage, e.g., \approach{} outperforms AFL++$_{Dict}$ by 26.2\% in \textit{strip}. Moreover, although AFL$_{Dict}$ and \qsym{} achieve higher edge coverage than \approach{} in \modifydata{\textit{xmlwf}, and \textit{size}} respectively, their performance gaps are rather limited, i.e., \modifydata{\approach{} underperforms AFL$_{Dict}$ in \textit{xmlwf} by 0.2\%, and \qsym{} in \textit{size} by 3.1\%}. Such results altogether indicate that \approach{} can achieve quite robust edge coverage performance. 

\subsubsection{Bug finding capability for bugs in the wild}

\begin{table}[htb]
    \centering
    \caption{The unique bugs explored by \approach{}}
    \label{tab:buglist}
    \begin{adjustbox}{width=0.95 \columnwidth}
    \begin{tabular}{|l|l|l|l|}
    \hline 
    \textbf{Project} & \textbf{Bug Type} & \textbf{Number}  & \textbf{Status}  \\
    \hline
    
    bison                           &       \Uouv{}             &   1   &   \unconfirmed{}      \\ \hline
    objdump                         &       \El{}               &   1   &   \fixed{}            \\ \hline
    bsdtar                          &       \Uouv{}             &   1   &   \unconfirmed{}      \\ \hline
    jasper                          &       \Af{}               &   1   &   \fixed{}            \\ \hline
    lou\_translation                &       \El{}               &   1   &   \fixed{}            \\ \hline
    libtiff                         &       \Uouv{}             &   7   &   \unconfirmed{}      \\ \hline
    objcopy                         &       \Uouv{}             &   1   &   \fixed{}            \\ \hline
    jhead                           &       \Uouv{}             &   6   &   \unconfirmed{}      \\ \hline
    precomp                         &       \badmalloc{}        &   1   &   \unconfirmed{}      \\ \hline
    nm                              &       \Ml{}               &   1   &   \confirmed{}        \\ \hline
   \multirow{2}{*}{zziplib}                           &       \Sbo{}              &   1   &   \unconfirmed{}      \\ 
                          &       \Sbo{}              &   1   &   \unconfirmed{}      \\ \hline
    \multirow{2}{*}{jpeginfo}       &       \Hbo{}              &   1   &   \fixed{}            \\ 
                                    &       \Uouv{}             &   1   &   \confirmed{}        \\ \hline    
    \multirow{3}{*}{cmix}           &       \mallocdismatch{}   &   1   &   \fixed{}            \\ 
                                    &       \memcpyoverlap{}    &   1   &   \fixed{}            \\ 
                                    &       \Uouv{}             &   1   &   \unconfirmed{}      \\ \hline      
    \multirow{6}{*}{bento4}         &       \Astb{}             &   1   &   \unconfirmed{}      \\          
                                    &       \Oom{}              &   2   &   \unconfirmed{}      \\          
                                    &       \Ml{}               &   1   &   \unconfirmed{}      \\          
                                    &       \Hbo{}              &   3   &   \unconfirmed{}      \\          
                                    &       \Sf{}               &   1   &   \unconfirmed{}      \\          
                                    &       \Huaf{}             &   1   &   \unconfirmed{}      \\ \hline   
    
    \end{tabular}
    \end{adjustbox}
    \end{table}
To evaluate the bug-finding capability, we apply \approach{} on our original benchmark suite and randomly select 10 additional real-world open-source projects (stars $>$ 100) from \git{} following prior evaluations~\cite{pata,redqueen,weizz,afl++}.
We also include all the grey-box fuzzers with \dictionarybased{} and \qsym{}, \meuzz{}, \pangolin{}, \angora{}, \redqueen{} in the evaluation of bug finding capacity.
To identify unique bugs, we first compile the selected projects with two additional sanitizers~\cite{ASAN, MSAN} to trigger crashes as possible. Next, we derive the unique crashes based on whether they incur unique execution paths following existing work~\cite{afl, afl++,aflfast,mopt,angora,klees2018evaluating,fairfuzz}. 
Finally, we manually analyze each unique crash to derive unique bugs. All bugs are categorized based on their root causes (their details are available in our \git{} page~\cite{githubrepo}). 
Table~\ref{tab:buglist} presents the unique bugs exposed by our approach. 
\approach{} has exposed \totalcrash{} previously unknown bugs where 30 of them cannot be exposed by other studied fuzzers within the given time limit. Moreover, \totalconfirmed{} of them have been confirmed and \totalfixed{} have been fixed by the corresponding developers. The results suggest that \approach{} is more effective than other fuzzers in terms of exposing real-world bugs. In particular, we list two examples of the exposed bugs below to illustrate the importance of the bugs found by \approach{}. We also demonstrate the details of all exposed bugs in our \git{} page~\cite{githubrepo} with their report links due to page limit.



\begin{figure}[htb]
    \lstset{style=demo-code}
    \centering
    \begin{lstlisting}[language=C, basicstyle={\fontsize{6.5}{6.5}\ttfamily}]
#define EXIF_JPEG_MARKER JPEG_APP0+1
#define EXIF_IDENT_STRING "Exif\000\000"
|\colorbox{backcolour}{[+] \#define EXIF\_IDENT\_STRING\_LEN 6}|
...
while (cmarker) {
    |\colorbox{uncoveredgray}{[-] if (cmarker->marker == EXIF\_JPEG\_MARKER) \;\;}|
    |\colorbox{backcolour}{[+] if (cmarker->marker == EXIF\_JPEG\_MARKER \&\&}|
    |\colorbox{backcolour}{cmarker->data\_length >= EXIF\_IDENT\_STRING\_LEN)}|
    if (!memcmp(cmarker->data,EXIF_IDENT_STRING,6)) 
        exif_marker = cmarker;
    cmarker = cmarker->next;
}    \end{lstlisting}
    \caption{A \hbo{} bug in \textit{jpeginfo}}
    \label{fig:jpeginfo-bug}
\end{figure}
\noindent \textbf{Heap-buffer-overflow bug in project \textit{jpeginfo}.} Figure~\ref{fig:jpeginfo-bug} shows the code snippet for a heap-buffer-overflow bug in project \textit{jpeginfo}, where in line 9, six bytes from \codeIn{EXIF\_IDENT\_STRING} are copied to \codeIn{cmarker->data} without any length checking, leading to a buffer overflow. 
To expose this bug, the \equal{} \codeIn{cmarker->marker == EXIF\_JPEG\_MARKER} should be satisfied. In our evaluation, only \approach{} reaches the branch guarded by such \equal{}. 
The developer fixed the bug~\cite{jpeginfo-bug-link} by adding buffer length checking statements from line 7 to line 8. They also replied to our bug report as follows:
\devcomment{``Thanks, looks like memcmp() may have read past end of the buffer in some circumstances.''}

\begin{figure}[htb]
    \lstset{style=demo-code}
    \centering
    \begin{lstlisting}[language=C, basicstyle={\fontsize{6.5}{6.5}\ttfamily}]
void ConvertUTF8(Word *W) {
    for (int I = W->Start; i < W->End; i++) {
        U8 c = W->Letters[i+1] + ((W->Letters[i+1]<0xA0)?0x60:0x40);
        if (W->Letters[I] == 0xC3 && (IsVowel(c) |$\|$| (W->Letters[i+1]&0xDF) == 0x87)) {
            W->Letters[i] = c;
            if (i+1 < W->End)
                |\colorbox{uncoveredgray}{[-] memcpy(\&W->Letters[i+1], \&W->Letters[i+2],}|
                |\colorbox{uncoveredgray}{W->End-i-1)}|; 
                |\colorbox{backcolour}{[+] memmove(\&W->Letters[i+1], \&W->Letters[i+2],}|
                    |\colorbox{backcolour}{W->End-i-1)}|;
            W->End--;
       }}}    \end{lstlisting}
    \caption{A memcpy-param-overlap bug in \textit{cmix}}
    \label{fig:cmix-bug}
\end{figure}

\noindent \textbf{Memcpy-param-overlap bug in project \textit{cmix}.} We have also reported a memory-overlapping bug in \textit{cmix} only exposed by \approach{}. The corresponding developers have fixed this bug after receiving our report~\cite{cmix-bug-link}. 
\remove{Figure~\ref{fig:cmix-bug} shows a memory-overlapping bug in \textit{cmix} where developers have accepted our fix suggestion after receiving our bug report~\cite{cmix-bug-link}. In line 7, the buggy code snippet copies the memory content from \codeIn{W->Letters[i+2]} to \codeIn{W->Letters[i+1]} via \codeIn{memcpy}. As the source and destination addresses differ by a single byte, it causes potential overlapped memory which is an undefined behavior in C/C++ programming. While the originally adopted function \codeIn{memcpy} fails to properly handle overlapping memory regions, applying the function \codeIn{memmove} can address such cases.}Figure~\ref{fig:cmix-bug} shows the corresponding buggy code snippet where in line 7, the memory content is copied from source \codeIn{W->Letters[i+2]} to sink \codeIn{W->Letters[i+1]} via \codeIn{memcpy}. As the source and sink addresses differ by a single byte, it causes potential overlapped memory, i.e., an undefined behavior in C/C++ programming. The function \codeIn{memcpy} does not guarantee proper handling of overlapping memory regions. In contrast, \codeIn{memmove} ensures accurate data replication in such cases. 
In our evaluation, only \approach{} generates seeds that expose this defect by satisfying the \equal{} in line 4, 
i.e., \codeIn{W->Letters[i]==0xC3} and \codeIn{IsVowel(c)}. 
The developers also commented on our report: 

\devcomment{``Thanks for the bug report, and the suggested fix! Changing to memmove fixed this.''}

\section{Threats to validity}
\noindent \textbf{Internal validity.} One threat to internal validity lies in the implementation of the studied techniques in our evaluation. To reduce this threat, we reused all the source code from the original projects directly in our implementation with best effort. When implementing the \dictionarybased{}, we proposed an automatic approach to extract tokens in a LLVM pass following prior work~\cite{fuzzingdriver} to reduce potential bias caused by user-provided tokens. 
Moreover, all the student authors manually reviewed the code of all studied fuzzers including \approach{} to ensure their correctness and consistency. 


\noindent \textbf{External validity.} The threat to external validity lies in the subjects and benchmarks. To reduce this threat, we have selected 9 representative state-of-the-art fuzzers which cover mainstream types of \assistfuzzings{}, including dictionary-based fuzzers, input-to-state-correspondence-based fuzzers, gradient-based fuzzers, and SMT-solver-based fuzzers. We also collect \totalbench{} frequently used projects from their original papers as our benchmark suite.  

\noindent \textbf{Construct validity.} The threat to construct validity mainly lies in the metrics used. To reduce this threat, we adopted the most popular metrics in fuzzing, i.e., edge coverage following~\cite{mopt, afl, afl++, fairfuzz, ankou}, to reflect the performance of different studied techniques. Furthermore, we evaluated the effectiveness of our approach in terms of the number of unique crashes.

\section{Related work}

\subsection{Fuzzing}
\label{sec:fuzzing}
Most existing fuzzers~\cite{fairfuzz, libfuzzer,aflfast, pata, wu2022evaluating, sjfuzz, jitfuzz, wu2023enhancing} use code coverage information to improve the efficiency of fuzzing. In particular, AFL~\cite{afl} provides the fundamental framework for coverage-guided fuzzers. Accordingly, Fioraldi et al.~\cite{afl++} integrated several techniques (e.g., taint tracking) to enhance the ability of exploring program states for AFL. She et al.~\cite{neuzz} proposed NEUZZ, which leverages the power of neural network models to explore unknown edges.\remove{, while Wu et al.~\cite{wu2022evaluating} further propose PreFuzz, a simple technique that enhances neural program-smoothing fuzzers with a resource-efficient edge selection mechanism.} Pham et al.~\cite{smartafl} proposed SGF which generates seeds on the virtual structure of the file rather than on the bit level to improve fuzzing efficacy. Meanwhile, \textit{AFLFast}  uses Markov Chain to schedule seeds for exploring program states~\cite{aflfast}. \fairfuzz{} focused on rare branches for its exploration~\cite{fairfuzz}. Zeror is a coverage-sensitive tracing and scheduling fuzzing framework that uses zero-overhead instrumentation and a schedule strategy between different instrumentation for AFL-based fuzzers~\cite{zeror}. Recently, researchers also pay attention to exploring program states by focusing on the diversity of the program behaviors. Nguyen et al.~\cite{bedivfuzz} introduced BeDivFuzz to schedule the mutation strategy towards the validity and diversity of program behaviors based on the received program feedback. Liang et al.~\cite{pata} proposed PATA to mutate the influencing input bytes by leveraging the power of diverse explored program paths. Yan et al.~\cite{pathafl} introduced a new approach PathAFL to reduce the tracing granularity of an execution path for exploring program states. QATest adopts a new coverage guidance and seed schedule strategy~\cite{qatest} for question-answering systems. EMS utilizes historical explorations to identify mutators that can trigger unique paths and crashes~\cite{ems}. 
By adopting an automatic dictionary generation strategy, \fuzzingdriver{} generates dictionary tokens for coverage-based grey-box fuzzers via parsing the original code~\cite{fuzzingdriver}. Compared to \fuzzingdriver{}, \approach{} schedules existing tokens in dictionaries for different seeds in runtime instead of generating tokens before fuzzing. 


To explore program states guarded by complicated constraints, hybrid fuzzing techniques are proposed to combine constraint solvers with grey-box fuzzers. Majumdar et al.~\cite{firsthybrid} presented hybrid fuzzing to interleave random fuzzing with constraint solver for deep exploration of program state space. Driller leverages fuzzing and selective concolic execution with constraint solver in a complementary manner to explore program states~\cite{driller}. Chen et al.~\cite{angora} leveraged the power of gradient descent to solve the constraints in the target program. \qsym{} optimistically solves constraints and prunes uninteresting basic blocks during fuzzing~\cite{qsym}. Chen et al.~\cite{meuzz} introduced a machine-learning-based seed scheduling strategy for hybrid fuzzing to explore program states efficiently. Huang et al.~\cite{pangolin} utilized polyhedral path abstraction to facilitate constraint solving. 
CONFETTI fuzzes Java programs by combining fuzzing with taint tracking and concolic execution with constraint solver~\cite{CONFETTI}. Meanwhile, \textit{chopped symbolic execution}~\cite{symbolicexecutionchop} leverages various on-demand static analyses at runtime to automatically exclude code fragments while resolving their side effects to improve the efficiency of constraint solving. Compared to constraint solver and dynamic taint tracking adopted by hybrid fuzzing, \approach{} generates a customized dictionary for each seed via a lightweight static analysis. Our experiments show that \approach{} outperforms the state-of-the-art constraint-solving fuzzers. 

\subsection{Studies on Fuzzing}
\label{sec:study}
Many empirical studies~\cite{fuzzbench, klees2018evaluating, havocdma, jiang2023evaluating, wu2022evaluating, gao2023vectorizing, zhao2022history, wu2020simulee} on fuzzing reveal various insights for improving fuzzing techniques. Donaldson et al.~\cite{gpufuzzing} investigated a variety of fuzzing techniques, including coverage-guided fuzzing with and without custom mutators to test compilers and processing tools for the graphics shading languages. Wu et al.~\cite{havocdma} conducted a study on \textit{Havoc} fuzzing strategy and demonstrated that it largely outperforms other strategies. 
B{\"o}hme et al.~\cite{reliability} performed a study on discussing the reliability metrics for evaluating the effectiveness of different coverage-based fuzzers. They also study the scalability 
 issues of fuzzing in vulnerability discovery~\cite{fuzzingcost}. 
FuzzBench is a open-source platform proposed for evaluating fuzzers to facilitate reliable and reproducible evaluation results~\cite{fuzzbench}. Herrera et al.~\cite{seedselection} systematically investigated and evaluated how seed selection affects a fuzzer's ability to expose vulnerabilities in real-world systems. Klees et al.~\cite{klees2018evaluating} provided multiple guidelines about how to evaluate the effectiveness of different fuzzers. In this paper, we conduct the first comprehensive study to investigate how \assistfuzzings{} perform in exploring program states and reveal various findings to facilitate future research. 
\section{conclusion}
In this paper, we investigated the strengths and limitations of \assistfuzzings{} for exploring program states. We first conduct an extensive evaluation to investigate how \assistfuzzings{} perform in exploring program states. The evaluation results suggest that \dictionarybased{} can be close to or even slightly more effective than other techniques. Next, we investigate their limitations and find that the \dictionarybased{} is most promising to be improved. Inspired by our findings, we present a lightweight approach namely \approach{} which customizes the dictionary for each seed. Our evaluation results show that \approach{} outperforms the best performer in our study by \modifydata{16.1\%} in terms of edge coverage. \approach{} also exposes \totalcrash{} previously unknown bugs where \totalconfirmed{} of them have been confirmed and \totalfixed{} of them have been fixed by the corresponding developers. 

\section*{Data Availability}
The data and code are available at \git{}~\cite{githubrepo} for public evaluation.
\section{Acknowledgement}
This work is partially supported by the National Natural Science Foundation of China (Grant No. 62372220) and Natural Sciences and Engineering Research Council of Canada (NSERC) Discovery Grant. It is also partially supported by the Leading Innovative and Entrepreneur Team Introduction Program of Zhejiang (Grant No. TD2019001) and Ant Group Research Fund.

\bibliographystyle{IEEEtran}
\bibliography{dictionary}
\end{document}